\documentclass[usenatbib]{mn2e}

\usepackage{graphicx}
\usepackage[normalem]{ulem}
\usepackage{color}
\usepackage{textcomp}
\usepackage{booktabs}
\usepackage{amssymb}
\usepackage[T1]{fontenc} 
\usepackage{aecompl}	
\usepackage{lastpage}   
\newcommand{\mnras}{MNRAS}

\def\apj{ApJ}

\def\apjs{ApJS}
\def\aj{AJ}
\def\aap{A\&A}

\def\mnras{MNRAS}

\def\pasp{PASP}

\newcommand{\revisedii}[1]{\textbf{#1}}

\newcommand{\Ha}{H$\alpha$}
\newcommand{\Hb}{H$\beta$}

\newcommand{\hii}{H~{\sc ii}}

\newcommand{\Nii}{[N~{\sc ii}] $\lambda$6584}
\newcommand{\nii}{[N~{\sc ii}]}

\newcommand{\Oiii}{[O~{\sc iii}] $\lambda$5007}

\newcommand{\oiii}{[O~{\sc iii}]}

\title[Retired galaxies and the star formation -- AGN connection]
{Retired galaxies: not to be forgotten in the quest of the star formation -- AGN connection}
\author[Stasi\'nska et al.] {
	  G. Stasi\'nska$^{1}$\thanks{E-mail:grazyna.stasinska@obspm.fr},
 	  M. V. Costa Duarte$^{1,2}$,
	  N. Vale Asari$^{3}$, R. Cid Fernandes$^{3}$, L. Sodr\'e Jr.$^{2}$\\
	   $^{1}$LUTH, Observatoire de Paris, CNRS, Universit\'e Paris Diderot; Place Jules Janssen 92190 Meudon, France\\
	   $^{2}$Departamento de Astronomia, Instituto de Astronomia, Geof\'isica e Ci\^encias Atmosf\'ericas, Universidade de S\~ao Paulo, S\~ao Paulo SP, Brazil \\
           $^{3}$Departamento de F\'{\i}sica - CFM - Universidade Federal de Santa Catarina, Florian\'opolis, SC, Brazil\\
	 \\
	}
\begin{document}

\date{Accepted .... Received ; in original form }

\pagerange{\pageref{firstpage}--\pageref{LastPage}} \pubyear{2014}

\maketitle

\label{firstpage}

\begin{abstract} 

We propose a fresh look at the Main Galaxy Sample of the Sloan Digital Sky Survey by packing the galaxies in stellar mass and  redshift bins. 
We show how important it is to consider the emission-line equivalent widths, in addition to the commonly used emission-line ratios, to properly identify retired galaxies (i.e. galaxies that have stopped forming stars and are ionized by their old stellar populations) and not mistake them for galaxies with low-level nuclear activity. 
We find that the proportion of star-forming galaxies decreases with decreasing redshift in each mass bin, while that of retired galaxies increases.  Galaxies with $M_\star > 10^{11.5} M_\odot$ have formed all their stars at redshift larger than 0.4.  The population of AGN hosts is never dominant for galaxy masses larger than $10^{10} M_\odot$. We warn about the effects of stacking galaxy spectra to discuss galaxy properties. We estimate the lifetimes of active galactic nuclei (AGN) relying entirely on demographic arguments --- i.e. without any assumption on the AGN radiative properties.  We find upper-limit lifetimes of about 1--5 Gyr for \emph{detectable} AGN in galaxies with masses between $10^{10}$--$10^{12} M_\odot$. The lifetimes of the AGN-\emph{dominated} phases are a few $10^8$ yr. Finally, we compare the star-formation histories of star-forming, AGN and retired galaxies as obtained by the spectral synthesis code {\sc starlight}. Once the AGN is turned on it inhibits star formation for the next $\sim$ 0.1 Gyr in galaxies with masses around $10^{10} M_\odot$, $\sim$ 1 Gyr in galaxies with masses around $10^{11} M_\odot$.  

\end{abstract}


\section{Introduction}

It is nowadays considered that episodes of nuclear activity are of prime importance in the evolution of galaxies and in the building up of the present-day Universe, due to the strong interplay believed to exist between the active galactic nuclei (AGN) and star formation (SF)  (see Fabian 2012 for a detailed review). Observational evidence of this interplay requires a large and complete sample of galaxies with adequate spectra and well defined criteria to classify galaxies into appropriate categories. One of the pioneering studies in this direction is that of Huchra \& Burg (1992), based on the CfA redshift survey (Huchra et al. 1983) which is a magnitude-limited spectroscopic survey of 2500 galaxies. It was found that classical Seyfert galaxies constitute about 1\% of field galaxies with absolute magnitudes $M_B$ smaller than $-20$,  and that the percentage of AGN hosts increases with luminosity, reaching 20\% for  galaxies with $M_B$ below $-21$ (using $H_0 = 100 \ {\rm km} \ {\rm s}^{-1}$). Since then, the Sloan Digital Survey (SDSS, York et al. 2000) has provided a spectroscopic database of nearly one million galaxies  which allows one to study the properties of galaxies up to  a redshift $z \sim 0.4$. Many studies have  used SDSS data to address the nature of AGN hosts  and the relation between nuclear activity and star formation (e.g. Kauffmann et al. 2003 -- referred to as K03 in this paper --; Schawinski et al. 2007; Lee et al. 2007; Deng et al. 2012; Lamassa \& Heckman 2013). Some studies have complemented optical data from the SDSS with data at other wavelengths (e.g. Kauffmann et al. 2007 for the UV; Constantin et al. 2009 for X-rays; Rosario et al. 2013 for the mid-infrared).

Most of these studies adopt the classification used by K03 based on the \Oiii/\Hb\ vs \Nii/\Ha\ emission-line ratio diagram introduced  by Baldwin et al. (1981, commonly referred to as the BPT diagram), in which galaxies are divided into star-forming (those below the divisory line introduced by K03),  ``pure'' AGN (those lying above the Kewley et al. 2001 -- K01-- divisory line), and the ``transition'' or ``composite'' ones lying in-between. As emphasized by Cid Fernandes et al. (2010, 2011) this classification leaves aside a large number of galaxies that cannot be classified in the BPT diagram because they lack some of the required emission lines in their spectra, most often \Hb. In addition, it does not account for the presence of ``retired'' galaxies (following the nomenclature introduced by Stasi\'nska et al. 2008). These systems have stopped forming stars and are ionized by their old stellar populations (namely the hot low-mass evolved stars, or HOLMES), without a detectable contribution of an AGN. Finally, the K03 line to separate  star-forming galaxies from AGN hosts has been drawn empirically without physical justification. Consequently, it is uncertain in the region of the BPT diagram where the SF and AGN wings merge.  Stasi\'nska et al. (2006, hereafter S06) showed that many galaxies below the K03 line do in fact contain an AGN, and proposed a divisory line based on photoionization models to separate \emph{pure} SF galaxies from galaxies containing a \emph{detectable} AGN. It is to be noted that the new photoionization models of Dopita et al. (2013) are in agreement with the S06 ones as regards the identification of pure SF galaxies. 

Cid Fernandes et al. (2011) proposed a different classification of galaxies which takes into account the existence of retired galaxies by considering the equivalent width of the \Ha\ line (in their WHAN diagram). In this paper, we show that the WHAN diagram leads to a much more comprehensive and reliable view of galaxy evolution than the commonly used BPT diagram -- although it also shares some of the drawbacks of the BPT diagram.
 
To properly address the problem of the star formation-AGN connection, it is important not only to use safe diagnostics of the nature of galaxies, but also to divide galaxies according to their  masses, since it known that mass is a fundamental parameter for the pace of galaxy evolution (Cowie et al. 1996; Heavens et al. 2004; Cid Fernandes et al. 2007; Asari et al. 2007; Jimenez et al. 2007; Haines et al. 2007). 

Another aspect not to be neglected is that past episodes of nuclear activity cannot be detected -- one can only recognize those that take place at the time when a galaxy is observed. Splitting the observational data in redshift slices therefore can provide some clues on the past activity of galaxies.  Other factors such as  morphology or environment also play a role in the evolution of galaxies and have been addressed by various authors (e.g. Haines et al. 2007; Bamford et al. 2009; Lietzen et al. 2011; Schawinski et al. 2010; 2012, 2014), and will need to be reexamined in future papers in the context of the present work. Here, we focus on just the following aspects: the correct assignation of galaxy spectral classes implying a proper identification of the retired galaxies, and the importance of presenting the results of any analysis in bins of mass and redshift.

Dividing the observational data into redshift bins has another important merit. It allows one to compare in each bin spectra encompassing regions of similar physical sizes and to study aperture effects by comparing bins of different redshifts. This is important since the 3\arcsec\ diameter SDSS fiber does not cover the entire galaxies, especially at low redshifts. Aperture effects are expected to be important  particularly for spirals, for which the central zones contain an old bulge and -- possibly -- an active nucleus while the outer zones correspond to the star-forming disk.
A full assessment of aperture effects requires  integral field spectroscopy of representative samples of  galaxies. A few preliminary studies have been published (Brinchman et al. 2004; Kewley, Jansen \& Geller 2005) but investigations based on integral field spectroscopy are just starting  (Gerssen et al. 2012; Mast et al. 2013; Iglesias-P\'aramo et al. 2013).  As will become evident in the course of the paper, for most aspects of our study we need not separate aperture affects and evolution in the interpretation of our results.

This paper is organized as follows. In  Sect. \ref{data}, we define our master sample of galaxies, explain the preliminary treatment applied to their spectra and divide galaxies in mass and redshift bins. In Sect. \ref{emlinediag} we discuss the BPT and WHAN emission line diagnostics including aperture effects and define  BPT and WHAN subsamples. 
In Sect. \ref{censusSDSS}, we provide a detailed census of the galaxy emission-line spectral types in our master sample. 
This allows us to proceed, in Sect. \ref{lifetimes}, to a direct estimation of  AGN lifetimes. In Sect. \ref{SFH}, we proceed one step further and analyze the star formation histories of the different categories of galaxies in mass and redshift bins, in an attempt to better understand the interplay between the star-forming, AGN and retired phases of galaxy evolution in the past 3--4 Gyr.  The main outcomes of the this study are summarised in Sect. \ref{summary}.

Three appendices complete this paper: Appendix \ref{app:sel} discusses selection effects in the SDSS Main Galaxy sample, Appendix \ref{app:stacks} warns about the effects of stacking galaxy spectra to discuss galaxy properties, and Appendix \ref{add} provides some additional material.

 Throughout the paper we consider a $\Lambda$CDM cosmology with 
$H_0 = 70 \ {\rm km} \ {\rm s}^{-1} \ {\rm Mpc}^{-1}$, $\Omega_m=0.30$,
and $\Omega_{\Lambda}=0.70$.

\section{The database}
\label{data}

\subsection{Sample selection}
\label{sample}

We consider the 7th Data Release of the Sloan Digital Sky Survey  \citep[SDSS/DR7,][]{Abazajianetal2009} which covers 9380 sq.deg. in the sky and presents fiber-fed spectroscopy in the wavelength range 3800--9200 \AA\  with mean spectral resolution of $\lambda/ \Delta\lambda\sim$1800 for nearly one million galaxies.  Since we are interested in the relative populations of different kinds of galaxies, we need  a well-defined  subsample. We therefore restrict ourselves to the Main Galaxy Sample, which is a complete flux-limited sample of galaxies down to a magnitude $m_r=17.77$   (Strauss et al. 2002). Note that, because we work with mass-redshift bins, we do not need a volume-limited sample, which would severely restrict the usable redshift range. This issue will be further discussed below.

We select galaxies with $z$-band covering factor larger than 20\% to reduce aperture effects in the determination of the stellar masses. We further impose a cut in redshift ($z > 0.002$) to garantee that luminosity distances are not dominated by peculiar motions (e.g. Ekholm et al. 2001). 
This reduces our sample to a total of 574,473 objects. 
Additional restrictions have to be made, depending on the properties under discussion, and will be explicited below.

\subsection{Analysis of the data}
\label{analysis}

Our work on the galaxy spectra makes use of the spectral synthesis code {\sc starlight} (Cid Fernandes et al. 2005). This is a inverse stellar population synthesis code which decomposes the observed spectra in contributions from simple stellar populations (SSPs) of given age and metallicity. The library of SSPs is based on the Bruzual \& Charlot (2003) evolutionary models of galaxies, adopting a Chabrier (2003) initial mass function (IMF), ``Padova 1994'' evolutionary tracks (Bertelli et al. 1994) and the STELIB spectral library (Le Borgne et al. 2003), as explained in detail in Asari et al. (2007). Spectral regions containing bad pixels, emission lines and the Na D doublet are not considered for the fits. The stellar extinction is an outcome of the {\sc starlight} fitting. It is computed using a Cardelli, Clayton \& Mathis (1989) extinction law with $R_V = 3.1$. The total masses of stars present in the galaxies, $M_\star$, are  obtained after correcting for the fraction of the galaxy $z$ band luminosity outside the SDSS fiber as explained in Cid Fernandes et al. (2005). The intensities of the emission lines were measured after subtracting the modelled stellar spectrum from the observed one. More details can be found in Mateus et al.  (2006), Stasi\'nska et al. (2006), Asari et al. (2007) and Cid Fernandes et al. (2010).  All the data used in this paper can be retrieved from the {\sc starlight} database\footnote[1]{http://casjobs.starlight.ufsc.br/casjobs/}.

We generally did not make any restriction concerning the signal-to-noise (S/N) ratio in the continuum since we are mainly interested in the stellar masses, which are very robust outputs from {\sc starlight}: as shown in Cid Fernandes et al. (2005), the uncertainty in $M_\star$ is of 0.1 dex for a S/N ratio of 5 at 4000 \AA, and objects with smaller S/N represent only 1.5  per thousand of the objects in our sample (3 per thousand of the ones with redshift $z > 0.2$).  Concerning the line intensities, we may have to restrict the sample by imposing criteria related to the quality of emission lines, as detailed below.

\subsection{Mass and redshift bins}
\label{mass-redshift}

In order to convey a synthetic view of the realm of SDSS galaxies we divide our samples of galaxies in bins of mass and redshift. 
We consider mass bins of 0.5 dex for log $M_\star$\footnote{Throughout the paper $M_\star$ is expressed in units of solar masses.} between 8.5 and 12.5. 
Concerning redshift, we consider bins of $\Delta z = 0.05$ until $z = 0.40$. 
These redshift limits correspond to sampled galactic radii of  
 1.5 to 8.1 kpc for the adopted cosmology.
The widths of the adopted mass-redshift bins are a compromise between i) obtaining a significant number of objects in each bin to robustly define   mean properties 
and ii) defining a sufficient number of bins in the mass and redshift ranges considered.  Note that the errors in the individual masses of the galaxies are much smaller than the width of the mass bins.

\section{Emission-line diagnostics of the galaxies}
\label{emlinediag}

\subsection{Some generalities}
\label{general}

\begin{table*} 
\begin{center}
\begin{tabular}{clcccc}
\hline
\multicolumn{2}{l}{ }&\multicolumn{2}{c}{tiny coverage}&\multicolumn{2}{c}{total coverage}\\
\cmidrule(l){3-4}\cmidrule(l){5-6}
             &         case                    		&   BPT 	&   WHAN 	&   BPT 	& WHAN 		\\
\hline

$a$          &  old bulge + SF disk				& L			&	R		&	C/SF 	&	AGN/SF	\\
$b$          &  AGN + SF disk					& AGN		&	AGN		&	C/SF 	&	AGN/SF	\\
$c$          &  AGN + SF bulge + SF disk		& C			&	AGN		&	C	 	&	AGN/SF 	\\
$d$         &  SF bulge + SF disk				& SF		&	SF		&	SF	 	&	SF		\\
$e$         &  old bulge + old disk/spheroid 	& L			&	R		&	L 		&	R		\\
\hline
\end{tabular}
\end{center}
\caption{Classification of galaxies using the BPT and WHAN diagrams, for two extreme fiber covering factors. ``C'' stands for ``composite, ``L'' stands for ``LINER'', ``R'' stands for ``retired''.}
\label{tab:contingency}
\end{table*}

As mentioned in the introduction, the vast majority of emission-line diagnostics of the ionization of galaxies are based on the BPT diagram where pure  SF galaxies are considered to lie below the K03 line and pure AGN above the K01 line, the space between the two lines being occupied by so-called composite objects.  Nevertheless, signs of AGN activity can be detected well below the K03 line, as shown by S06. On the other hand, ``pure'' AGNs lie well above the K01 line.  Note that Kewley et al. (2013)  now define another line to separate star-forming galaxies from galaxies containing an AGN, which actually is very close to the one of S06.

As shown by Stasi\'nska et al. (2008), the BPT diagram is however not able to distinguish between galaxies containing a weak AGN and retired galaxies, i.e. galaxies ionized by their  HOLMES (which can also explain the ionization of the extraplanar gas in spiral galaxies, see Flores-Fajardo et  al. 2011). The existence of such galaxies has  since then been evidenced by integral-field spectroscopy of galaxies, showing that the extended  emission seen in low-ionization galaxies is inconsistent with a central point source for the ionization, ruling out nuclear activity as the dominant source for the emission (Sarzi et al. 2010; Kehrig et al. 2012; Papaderos et al. 2013; Singh et al. 2013).
Cid Fernandes et al. (2011) proposed to use the equivalent width of H$\alpha$, EW(\Ha), as a function of \nii/\Ha\  (the WHAN diagram) to distinguish between galaxies containing an AGN and retired galaxies. This is the second diagnostic diagram we will consider here. Another advantage of using the WHAN diagram instead of the BPT is that it requires only 2 lines, while the BPT diagram requires 4 lines. We must note, however, that neither the BPT diagram nor the WHAN diagram are able to identify AGNs  in low-metallicity galaxies (S06; Groves et al. 2006).

The BPT diagram  can also be used to rank the metallicities in the star-forming wing, since the \oiii/\nii\ ratio has been shown since Alloin et al. (1979) to be a metallicity indicator. The \nii/\Ha\ ratio has also been shown to be strongly correlated with metallicity (Storchi-Bergmann et al. 1994; Van Zee et al. 1998) in giant \hii\ regions, and can thus serve to  rank metallicities in the SF region of the WHAN diagram. In the AGN/retired zone, however, its interpretation is not straightforward since \nii/\Ha\  both depends on the heating power of the main ionizing source (AGN or HOLMES) and on the nitrogen enrichment.

\subsection{Aperture effects on the BPT and WHAN diagrams}
\label{aperture}

 An important thing to realize is that both the BPT  and the WHAN diagrams are sensitive to aperture effects. More explicitly, the galaxy classification inferred from these diagrams depends on the relative proportions of bulge (or spheroid) and disk  sampled by the SDSS fiber.

The contingency table  presented in Table \ref{tab:contingency} schematically shows the diagnoses that would be obtained from the BPT diagram in its classical use (since Kewley et al. 2006) and from the WHAN diagram for the five possible galaxy configurations and two extreme fiber covering factors. We note that in case $a$,  both the BPT and the WHAN diagram may interpret the emission-line pattern as being a consequence of the presence of an AGN (since ``composite'' means AGN + star-forming) while the galaxy contains no AGN at all! The BPT, in addition, finds three more instances of the presence of an AGN (in terms of LINER) while the galaxy contains no AGN at all. These are cases $a$ and $b$ for a tiny fiber coverage, and case $e$ for a total fiber coverage.

\subsection{The BPT and WHAN subsamples}
\label{BPT-WHAN_sub}

From our parent sample, we define two subsamples: a sample based on the WHAN diagram (sample W), and a subsample based on the BPT diagram (sample B).
Sample W is the least restrictive of the two.
 Since within the WHAN diagram there is a continuity between emission-line galaxies and lineless galaxies in the sense that lineless galaxies would be found at vanishing values of EW(\Ha), here we do not impose a condition on S/N on the line intensities. This, of course, results in a somewhat uncertain classification of galaxies with low S/N but  should not bias the global picture. On the other hand, we remove galaxies where defects in the spectra at the wavelengths of \Ha\ and \nii\ do not allow even a coarse evaluation of the ratio of their intensities. To do this, we do not consider the criterion of Cid Fernandes et al. (2010), since we judge it to be too severe. Instead, we  measure the nebular velocity dispersion $\sigma_\mathrm {gas}$ for the \Ha\ and \nii\ lines and count the pixels within $\pm 1 \, \sigma_\mathrm{gas}$ of the peak of each line.  We then remove from our sample galaxies with more than 25\% bad pixels on either emission line.  The total number of galaxies that were removed with this criterion is 64,802, leaving 509,671 galaxies in sample W. 
Sample B contains 217,391 galaxies, which are selected by applying to sample W the  requirement that the S/N must be larger than 3 for all the four BPT diagnostic lines: \oiii5007, \Hb, \nii6584 and \Ha.   

\clearpage

\subsection{The BPT and WHAN diagrams in mass and redshift bins}
\label{BPT-WHAN-m-z}

\begin{figure}
\centering
\includegraphics[scale=0.53, trim=5mm 0 0 0, clip]{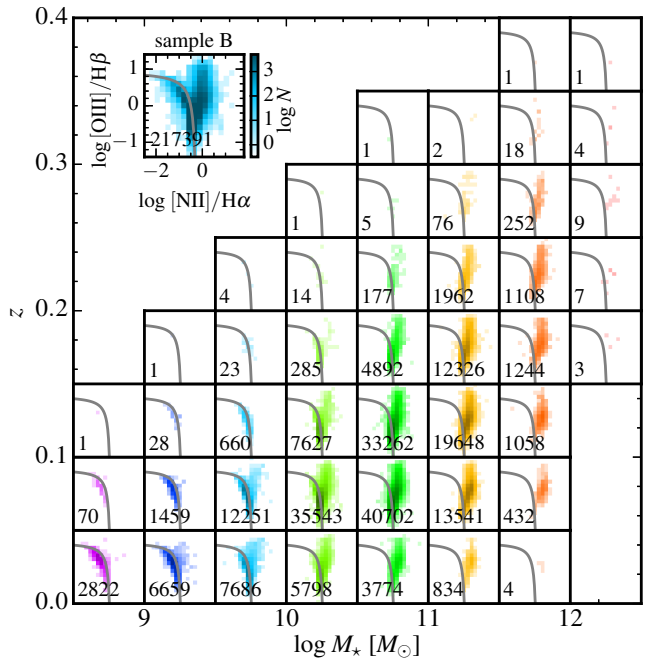}
\caption{\revisedii{BPT diagrams for sample B (i.e., a subset of sample W of objects with good S/N in all four BPT lines) in  mass-redshift bins. The number of objects contained in each panel is indicated. The color scale is in logarithm of the number of objects per pixel, as indicated in the color bar in the inset which shows the entire sample B in the BPT diagram. The curves are the S06 delimitations between pure SF galaxies and galaxies containing an AGN.}}
\label{BPT}
\end{figure}

Fig.\ \ref{BPT} show the BPT diagram for sample B in  the $(M_\star, z)$ bins we have defined. The total number of objects in each bin is indicated. The curves represent the S06 line to distinguish pure SF galaxies from AGN hosts.  As expected the lowest mass bins are populated only for the lowest redshifts (because the sample is flux limited), and the highest mass bins have very few objects below $z = 0.1$ (because of the decrease of the galaxy mass function towards high masses, see e.g. Panter et al. 2004). We clearly see how -- at any redshift -- the right wing becomes more and more populated with increasing stellar mass, leaving virtually no object in the SF wing at the highest masses.

\begin{figure}
\centering
\includegraphics[scale=0.53, trim=5mm 0 0 0, clip]{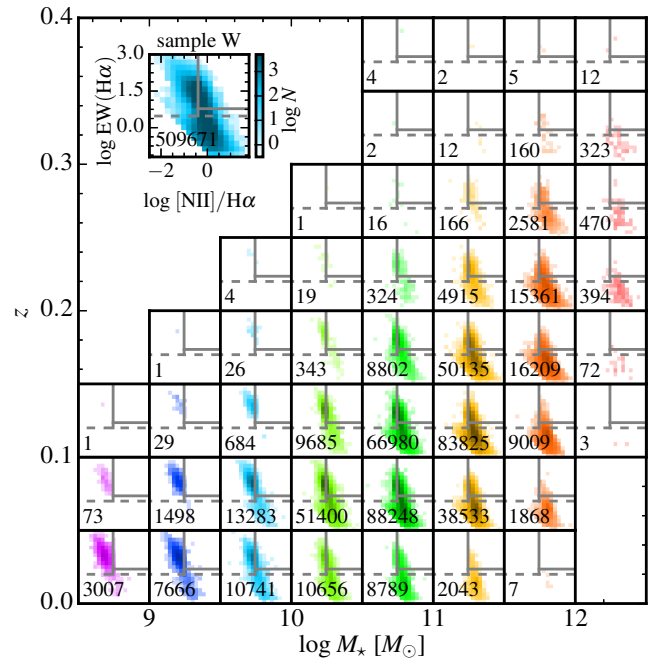}
\caption{\revisedii{WHAN diagrams for sample W (i.e., only removing galaxies with bad pixels in \nii6594 or \Ha) in  mass-redshift bins. The number of objects contained in each panel is indicated. The color scale is in logarithm of the number of objects per pixel, as indicated in the color bar in the inset which shows the entire sample W in the WHAN diagram. The continuous grey  lines indicate the delimitations between SF galaxies, galaxies with strong and weak AGNs, and the dashed grey line marks the limit for retired  galaxies according to Cid Fernandes et al. (2011).}}
\label{WHAN}
\end{figure}

Fig.\ \ref{WHAN} shows the WHAN diagram for sample W in  the  $(M_\star, z)$ bins.   The  grey  lines indicate the delimitations between the four categories of emission-line galaxies according to Cid Fernandes et al. (2011). In this diagram we also see how the SF zone becomes gradually depopulated with increasing stellar mass. But here we understand that this change essentially benefits the category of retired galaxies and not to that of  AGN hosts. 


\section{A detailed census of galaxy emission-line spectral types from the SDSS data}
\label{censusSDSS}

Cid Fernandes et al. (2010) discussed the proportions of pure star-forming galaxies/AGN hosts/retired galaxies for a volume-limited sample of SDSS galaxies with emission lines without paying special attention to their masses and redshifts. Here, we extend this discussion by evaluating these proportions in mass and redshift bins.

The first thing to consider in such an endeavour is the question of completeness. If observational selection removes some kind of galaxies from our sample, we must take this into account in the overall census. By working in  ($M_\star, z$) bins, we are less prone to discriminations against certain categories of objects (defined by their colours or by their emission-line properies). 
In Appendix \ref{app:sel} we discuss in detail  the bias expected in each ($M_\star, z$) bin. It turns out that some bins are definitely free from colour bias. At the smallest redshifts, the bins with log $M_\star > 10$ are complete.  In the next redshift interval,  bins  with log $M_\star > 10.5$  are complete, and so on.  All the subpanels where the sample is complete  are flagged with a  yellow background  in Fig. \ref{sel_WHAN}. As shown in Appendix \ref{app:sel}, adjacent bins are not much affected  by colour bias. As redshift increases, the limiting magnitude $m_r$ of  the Main Galaxy Sample will  reduce the number of complete mass bins by playing against galaxies of  smaller masses.  What is problematic for our study is that there could be a colour discrimination against some kind of galaxies. 
A general property of a survey limited by magnitude in the $r$ band is that the galaxies that are more easily missed are \emph{red} ones, since blue galaxies with same masses are more luminous, as seen in Appendix \ref{app:sel}. Thus, at the lowest redshift,  the  log $M_\star < 9$ bin could be missing an important number of red galaxies. However, the selection bias will become less important for $9 < $log $M_\star <9.5$, and will probably vanish for log $M_\star > 10$. In the next redshift bin, the selection bias will start being noticeable for log $M_\star < 10.5$, an so on.

\subsection{BPT-based demography}


\begin{figure*}
\centering
\includegraphics[scale=0.85, trim=10 350 10 70, clip]{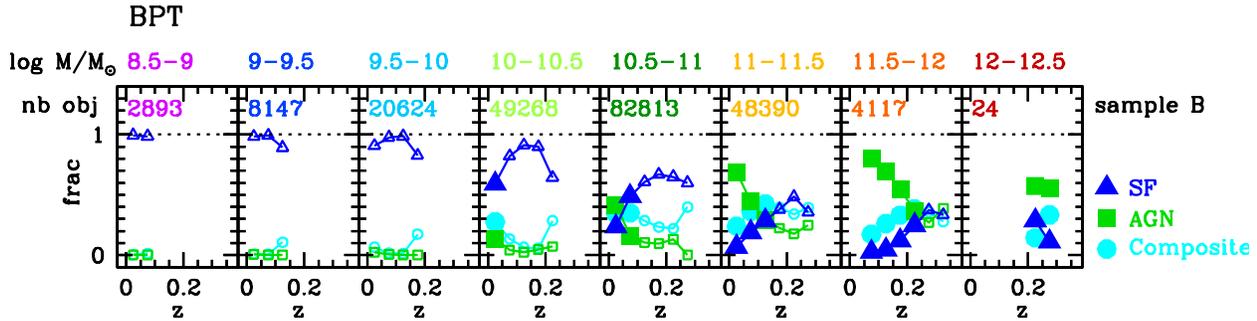}
\caption{Fraction of galaxies of different emission-line spectral types as a function of redshift for the different mass bins. The spectral types are  obtained from the canonical use of the BPT diagram.  Blue triangles: SF; green squares: AGN;  cyan circles: composite. Large symbols correspond to ($M_\star, z$) bins that are judged devoid of colour bias (see text and  Appendix \ref{app:sel}). }
 \label{fraction-BPT-redshift}
\end{figure*}

Before turning to the WHAN diagnostic diagram, it is instructive to first carry out a  demographic study using the BPT diagram in the way promoted by K03, since this is still the most popular  way to separate galaxies into star-forming and AGN hosts.

Figure \ref{fraction-BPT-redshift} shows the fraction of galaxies of different emission-line spectral types \revisedii{for sample B} as a function of redshift for our different mass bins, as obtained from the canonical use of the BPT diagram, i.e. using the K03 and K01 lines to distinguish  between SF, ``composite'' and AGN  galaxies.
In each panel we indicate the total number of objects represented.  The larger symbols represent ($M_\star, z$) bins that are judged free of bias  (see Appendix  \ref{app:sel}), and adjacent bins are probably little affected by bias. 
\revisedii{We only plot bins with a minimum of 5 galaxies.}
We see that for log $M_\star < 9$ all the galaxies are of SF type; AGN hosts are found only for masses larger than that.  Until log $M_\star = 10$, SF types still constitute the dominant population of galaxies. However, as noted above, the Main Galaxy Sample may be missing a population of low-mass red galaxies, whose importance is difficult to assess. Among this population, galaxies containing an AGN but not experiencing present-day star formation could perhaps exist.   At higher values of $M_\star$ the tendency starts reversing and  for log $M_\star > 11.5 $ we find that the population of galaxies is vastly dominated by AGN hosts. We also note that the proportion of SF galaxies always tends to decrease with decreasing redshift while the proportion of AGNs increases. Qualitatively, this is what is expected from mere aperture effects where the contribution of the old bulge with respect to the star-forming disk increases as redshift decreases.

\subsection{WHAN-based demography}


\begin{figure*} 
\centering
\includegraphics[scale=0.85, trim=10 270 10 70, clip]{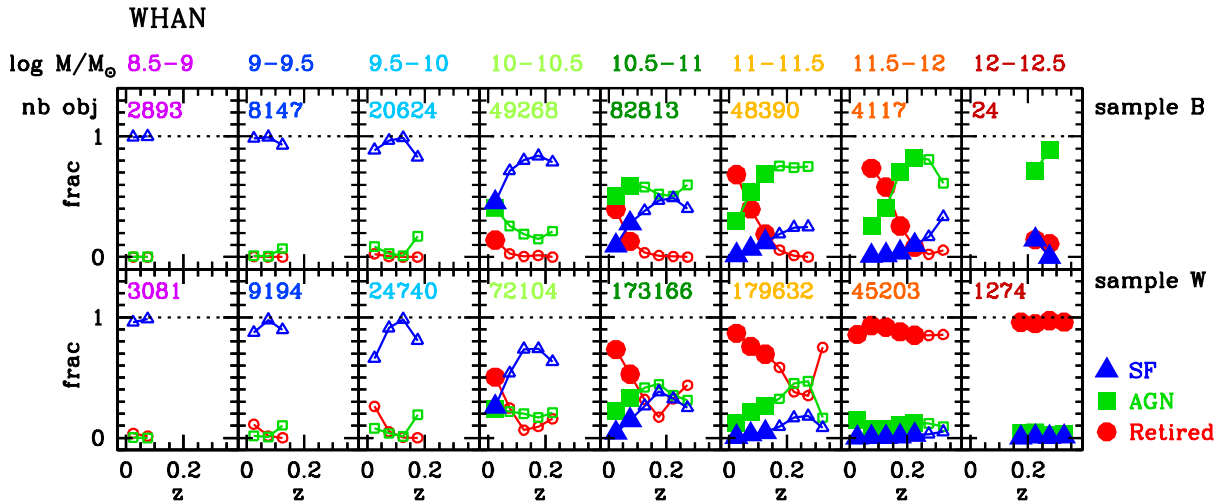}
\caption{Fraction of galaxies of different emission-line spectral types (blue triangles: SF; green squares : AGN (strong and weak); red circles: retired (with or without emission lines) as a function of redshift for the different mass bins as obtained from the  WHAN diagram.  Large symbols correspond to ($M_\star, z$) bins that are judged devoid of colour bias. \emph{Top:} WHAN classification of sample B, which imposes a S/N limit to \oiii5007, \Hb, \nii6584, and \Ha. \emph{Bottom:} WHAN classification of sample W, which only excludes galaxies whose spectra around \nii6584 or \Ha\ is unreliable.
}
 \label{fraction-WHAN_all-redshift}
\end{figure*}

As commented before, the BPT diagram demands four lines to be observed with good S/N, which selects against weak-line objects and leaves aside about half of the galaxies of our master sample. More importantly, the BPT diagram is not able to distinguish between hosts of weak AGN and retired galaxies since both categories display similar line ratios. Emission-line equivalent widths need to be considered as well.

Now we classify the galaxies according to the WHAN diagram, which overcomes these limitations of the BPT.
Figure \ref{fraction-WHAN_all-redshift} shows the fraction of SF galaxies, AGN hosts (putting strong and weak AGN in a single category) and retired galaxies (with or without emission lines) as a function of $z$ in the different mass bins in our sample,  for both samples B and W.

Let us first focus on the differences between Fig. \ref{fraction-BPT-redshift} and the top panel of Fig. \ref{fraction-WHAN_all-redshift}. Both sets of plots show galaxies in mass and redshifts bins for sample B, the former classifying galaxies based on their position on the BPT and the latter on the WHAN diagram. For masses $10.5 < \log M_\star < 12$, the BPT classes tricks us into finding an increase in AGN activity at smaller redshifts. When classifying \textit{ the same} set of galaxies using the WHAN this trend disappears. By telling apart AGN hosts from retired galaxies, the staggering conclusion is that a great part of what is attributed to AGN behaviour on the BPT is in fact just due to galaxy retirement.
Given the dispersion of galaxy types within a mass-redshift bin, one needs to be extremely careful when working with averaged galaxy properties or stacked spectra (see Appendix \ref{app:stacks}), even when binning in stellar mass and redshift.

The bottom panel of Fig. \ref{fraction-WHAN_all-redshift} shows galaxies as classified on the WHAN diagram for sample W, which is much more inclusive than sample B.

The general behaviour is that, as redshift decreases, in each mass bin the proportion of SF galaxies decreases, the proportion of retired galaxies increases while the proportion of AGN hosts rather tends to decrease (at least for  log $M_\star > 10.5$). 
 As was the case  in Fig. \ref{fraction-BPT-redshift}, we see that in the Main Galaxy Sample SF galaxies dominate the whole population of galaxies at any redshift for  log $M_\star < 10.5$. On the other hand, retired galaxies always dominate the whole population for   log $M_\star > 11.5$. In the intermediate mass bins, retired galaxies dominate at the lowest redshifts. As was the case for the BPT diagram, the redshift behaviour is qualitatively what is expected from aperture effects alone. For example, at intermediate masses the increasing proportion of retired galaxies as redshift decreases is expected as the fibre samples decreasing portions of the star-forming disk. 
This behaviour can also be due to galaxy evolution.

In spite of this uncertainty in the interpretation, in absence of any colour bias,  Fig. \ref{fraction-WHAN_all-redshift} would nicely depict the downsizing paradigm in the local Universe. The vast majority of galaxies with log $M_\star > 11.5$ have formed all their stars at redshift larger than 0.4. Galaxies with $10.5 < \log M_\star < 11.5$ gradually stop forming stars between $z=0.4$ and the present time. Galaxies with log $M_\star < 10.5$ still form stars presently. However, as explained in Appendix \ref{app:sel}, we expect the Main Galaxy Sample to miss red galaxies for masses below a certain threshold mass  to an extent that we are not able to evaluate, but which is certainly more important for the lowest masses. This means that the redshift evolution of the spectral types of galaxies with masses below $10^{10.5} M_\odot$\ cannot be obtained from the SDSS Main Galaxy Sample (although we will see in Sect. \ref{lifetimes} that our spectral-type census is probably correct down to $10^{10} M_\odot$).

In the mass and redshift bins that we consider free of colour bias, the population of AGN hosts is never predominant (except at the highest reshift of the $11 < \log M_\star < 11.5$ bin where they outnumber the retired galaxies by a tiny margin). Note that we reach  such a conclusion in spite of the fact that our definition of AGN host is less restrictive than the one which is generally used (i.e. that of K03). Also noteworthy is the fact that we do not see an increase in the proportion of AGN hosts with decreasing redshift, as was the case with the BPT diagram. Indeed, this increase is fake, and only due to retired galaxies being mistaken for AGN hosts.

We have found  that the standard use of the BPT diagram and our use of the WHAN diagram provide very different panoramas of the demographics of galaxy emission-line spectral types across the past 4 Gyr (which is the lookback time corresponding to  $z=0.4$). One advantage of the WHAN classification is that it does not set aside a large number of the galaxies. 
The other one is that it does not attribute to nuclear activity the emission-line ratios seen in weak-line objects. 
In the remaining of the paper, we therefore only consider the galaxy spectral classes that we defined with the help of the WHAN diagram (i.e. SF, AGN hosts and retired)
and only analyze sample W.

\section{AGN lifetimes}
\label{lifetimes}

\begin{figure*} 
\centering
\includegraphics[scale=0.7, trim={0 410 0 0mm}, clip]{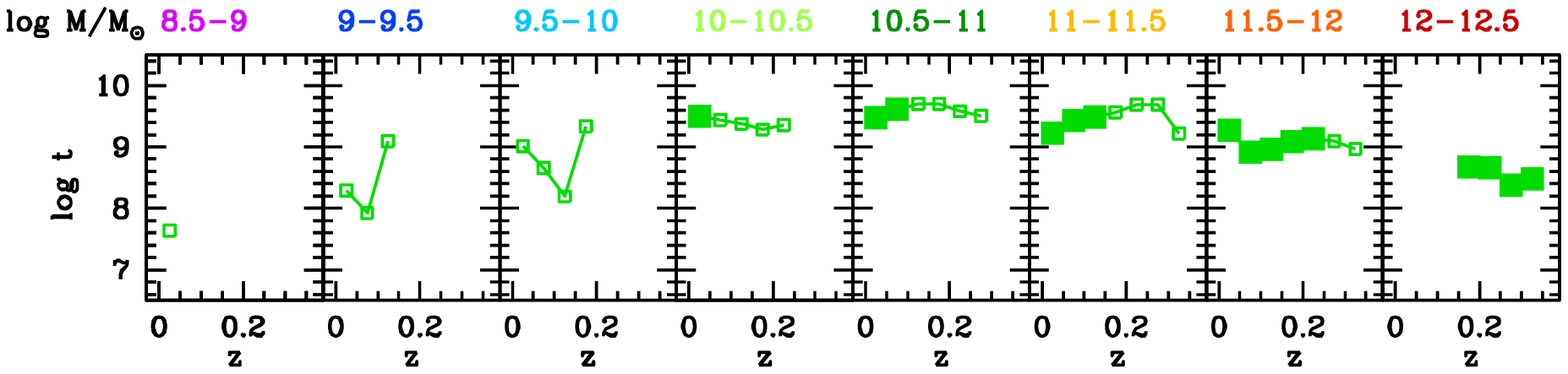}
\vspace {-8mm}
\includegraphics[scale=0.7, trim={0 410 0 0mm}, clip]{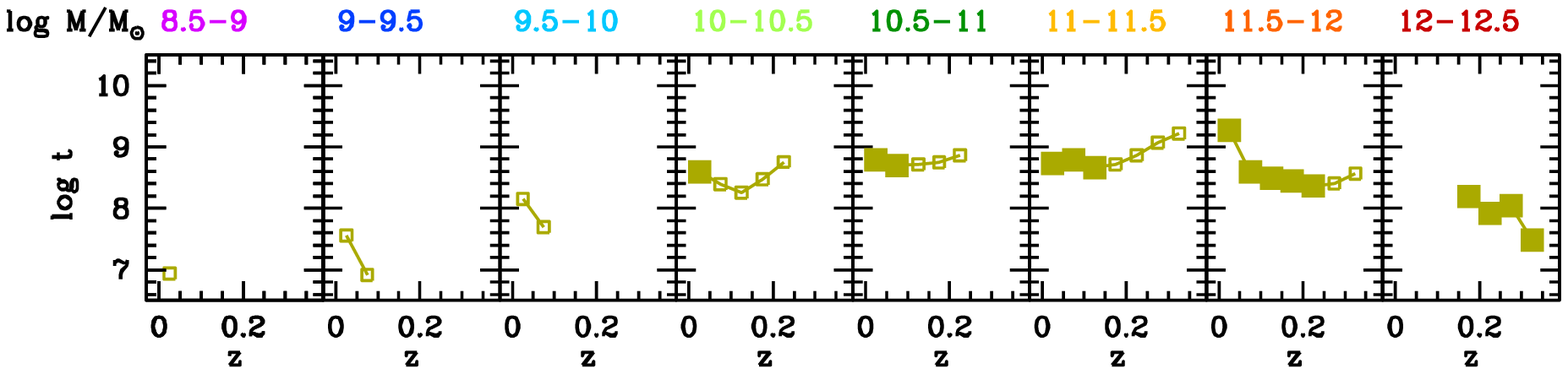}
\caption{AGN lifetimes computed from the census of AGN hosts in the  WHAN diagram as a function of redshift, for different mass bins. \emph{Top:} considering all galaxies with a \emph{detectable} AGN; \emph{Bottom:} considering AGN-\emph{dominated} galaxies. As in Fig. \ref{fraction-BPT-redshift},  large symbols correspond to ($M_\star, z$) bins that are judged devoid of colour bias.} \label{tAGN}
\end{figure*}

AGN lifetimes have been estimated by a number of methods based on black hole demographics or AGN radiative properties  which led to quite dispersed results (Martini 2004). Part of this dispersion may be due to the fact that the different studies do not actually refer to the same objects (e.g. high or low level of activity, massive or less massive galaxies). 

From the census of emission-line spectral types presented above, we can estimate the AGN lifetimes in each mass bin (and even in each redshift bin) by dividing the  number of AGN hosts  in a given bin by the total number of galaxies in that bin and multiplying it by the age of the Universe at the corresponding redshift.  This is shown in Fig. \ref{tAGN}. We find lifetimes of the order of 1-5 Gyr for log $M_\star$ between 10 and 12. Statistics for lower masses are affected by the colour bias discussed in Appendix  \ref{app:sel} so the lifetimes are not reliable: since we probably miss retired/red galaxies at a higher proportion in the biased bins, the fractions of AGNs and SF in such bins are probably upper limits. At higher redshift, the AGN population is also contaminated by bulge + disk systems in disguise (case $a$ of Table \ref{tab:contingency}). 
This means that the AGN lifetimes we show are also upper limits in these cases.  In the 12--12.5 mass range, the lifetimes appear to be shorter, about 0.3 Gyr.  We can do the same, now restricting ourselves to those objects where the AGN heavily contributes to the emission lines (i.e. galaxies which have EW(\Ha) $ > 3$\AA\ \emph{and} are above the K01 line; for simplicity we will refer to them AGN-\emph{dominated} galaxies, although this term is somewhat inaccurate since stellar ionization might be preponderant in a number of cases). This is done in the bottom panel of Fig. \ref{tAGN}. Now the lifetime appears to be of the order of a few $10^8$ yr. This implies that strong AGN are short-lived with respect to weak AGN, or that the phase of strong nuclear activity is shorter by a factor of 3--10 than the period where nuclear activity can be detected.

Note that for $10 < \log M_\star < 12$ we do not see any significant variation of AGN lifetimes with respect to redshift even in the redshift domain where absence of colour bias is not guaranteed.  This can perhaps be taken as  an indication that the colour bias is not very strong for those bins. 
Neither do we see any strong variation of the estimated AGN lifetimes with respect to $M_\star$. 

It must be noted that the  obtained lifetimes are strongly related to the criterion used to define the activity, as indeed illustrated by the top and bottom panels of Fig.~\ref{tAGN}, and care must be taken when they are used in other contexts.

\section{Star formation histories }
\label{SFH}

We have just seen in the previous section that galaxies hosting AGN are, on average, more massive than SF galaxies\footnote{This was already found by K03, even though their AGN sample was strongly contaminated by retired galaxies, which are on average more massive than SF ones.}. It is also well-known that  ``red and dead'' galaxies are more massive than galaxies forming stars presently (Heavens et al. 2004).  However, at a given mass and redshift, some galaxies have already stopped forming stars while others still form them, some galaxies host an AGN while others do not. 
It has repeatedly been shown by detailed studies of nearby objects that nuclear activity in Seyfert galaxies is  linked with some level of star-forming activity  (Shlosman 1990; Storchi-Bergman et al. 1996; Raiman \& Storchi Bergmann 2000; Veilleux 2001; Gonz{\'a}lez Delgado et al. 2009). 
In order to further discuss the relation between star formation and nuclear activity, here we study the star-formation histories of the various categories of galaxies considered.

\subsection{Determination of the star-formation histories}
\label{detsfh}  

The star-formation histories can be obtained from the {\sc starlight} analysis of the galaxy spectra, which provides a description of galaxies content in terms of simple stellar populations of different ages. The specific star-formation rate, SSFR($t$), which measures the mass converted into stars at time $t$ with respect to the total mass converted into stars, is computed for each galaxy by smoothing the stellar population age distribution,  as explained in Asari et al. (2007, their Eq. 6) with the following modification.  For a galaxy at a given redshift, part of the light (hence stellar mass) may be attributed by {\sc starlight} to stellar populations older than the age of the Universe at that redshift. In order to correct for that, we find the lookback time $t_\mathrm{max}$ corresponding to each galaxy redshift and consider that the SSFR($t$) at $t = t_\mathrm{max}$ is the sum of the contributions of all stellar populations older than $t_\mathrm{max}$.

Note that the time resolution of stellar population analyses decreases with lookback time, so that it is not possible, for any given galaxy, to detect short episodes of star formation that happened long ago. Besides, the history of the activity in galaxies is not recorded in their emission lines. 
A statistical way to circumvent these limitations is to follow a population of galaxies through different redshifts. This allows us not only to detect short episodes of star formation which are nowadays too old to be discerned, but also to directly measure the emission-line properties in this past epoch.
Our dataset allows us to peep into the starburst-AGN connection in the last 4 Gyr of the life of the galaxies (with the caveat that we look at increasingly large portions of galaxies as redshift increases).

\subsection{Star formation histories of the various galaxy spectral types}
\label{sfhindiv}

 \begin{figure}
\centering
\vspace {0mm}
\includegraphics[scale=0.38, trim={0 0 0 10mm}, clip]{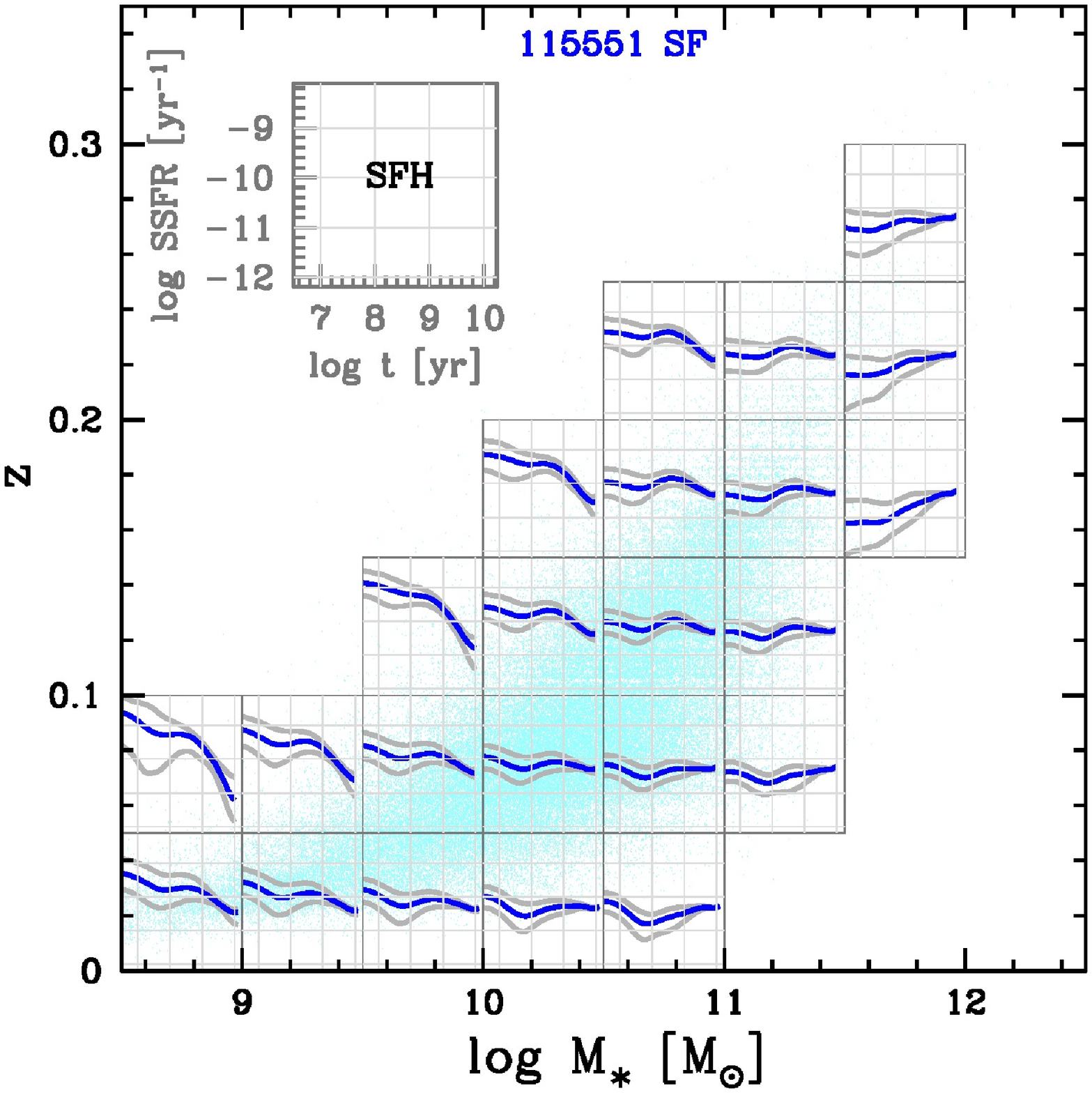}
\vspace {-0mm}
\includegraphics[scale=0.38, trim={0 0 0 10mm}, clip]{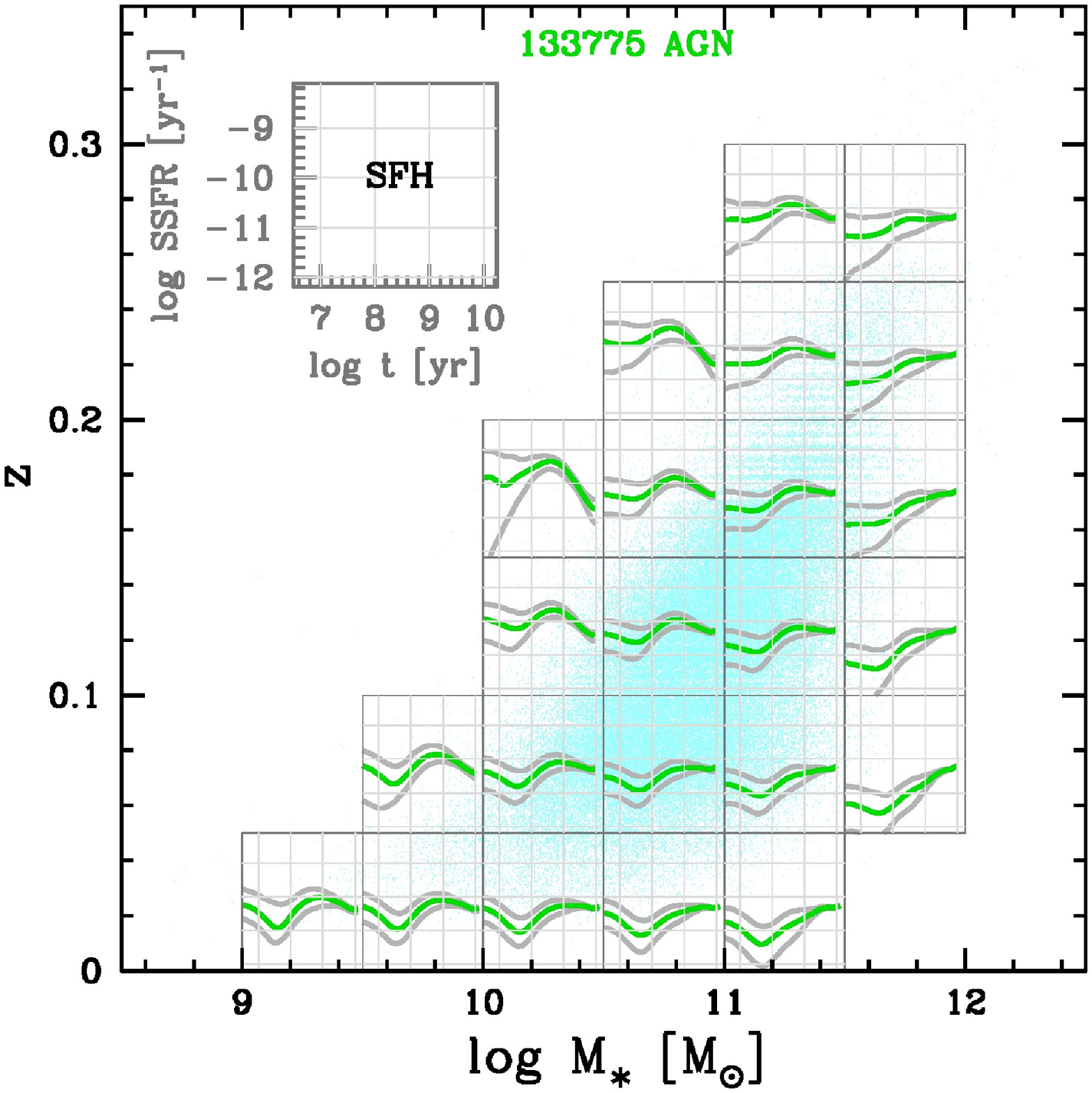}
\vspace {-0mm}
\includegraphics[scale=0.38, trim={0 0 0 10mm}, clip]{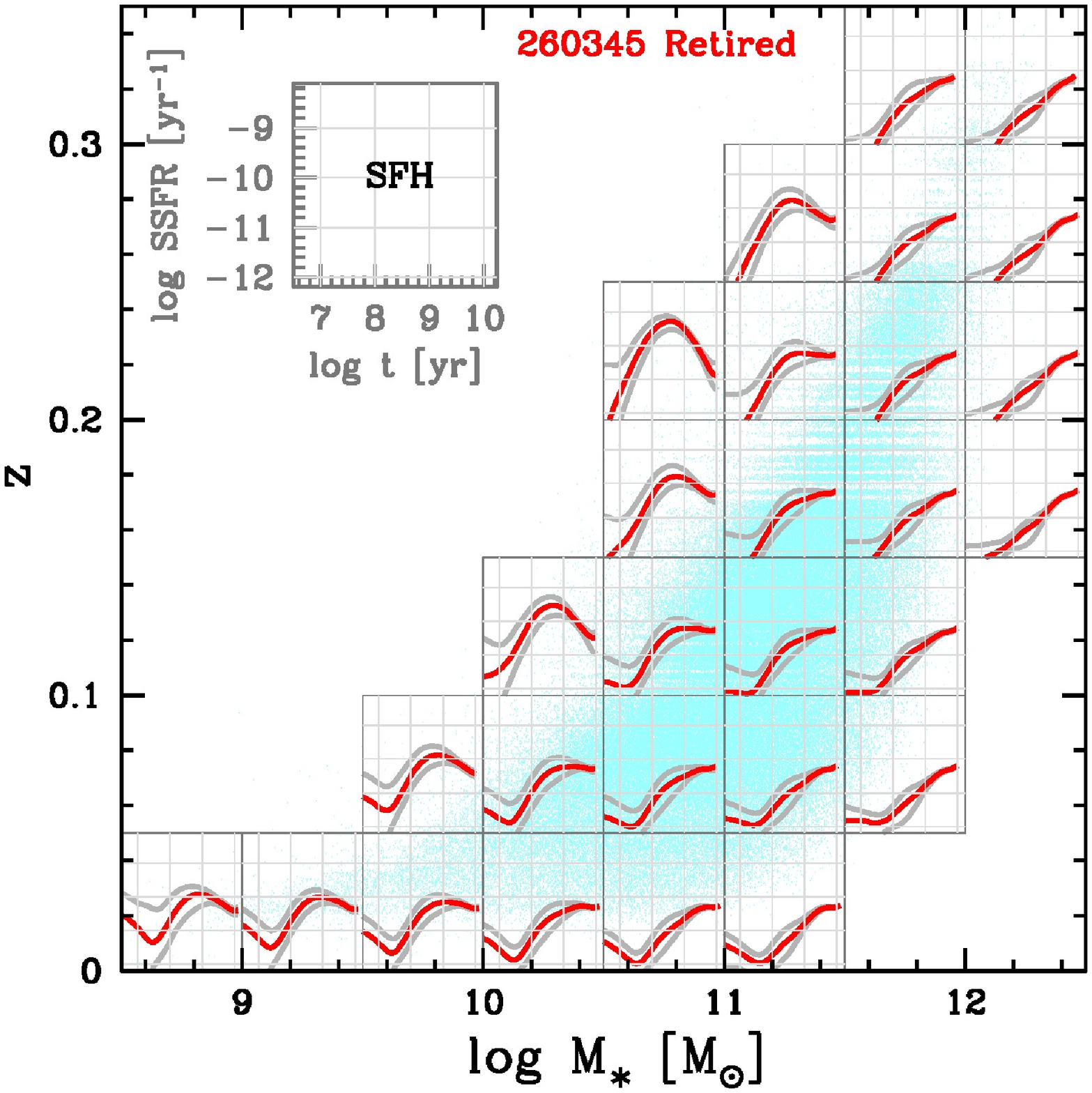}
\vspace {-2mm}
\caption{Star-formation histories, i.e. SSFR vs lookback time in the galaxies redshift frame  in  ($M_\star, z$) bins.  \emph{Top:} SF galaxies;  \emph{middle:} AGN hosts; \emph{bottom:} retired galaxies. Thick curves: median SSFR; thin curves: 16th and 84th percentiles. SSFRs are  shown only for bins with at least 50 objects. The cyan background points show the ($M_\star, z$) diagram for the relevant subsample of galaxies.}
 \label{SFH-indiv}
\end{figure}

Figure \ref{SFH-indiv} shows the variations of the median SSFRs (thick blue curves) and 16th and 84th percentiles (thin curves) as a function of lookback time in each of our mass-redshift bins for our three galaxy spectral types. The top panel concerns SF galaxies, the middle panel concerns AGN hosts, and the bottom panel concerns retired galaxies. Only bins  containing at least 50 objects are shown, to allow for a robust representation of the star-formation history in the considered bins. For convenience, in each ($M_\star, z$) subpanel, we consider the lookback time with respect to the galaxy redshift, and not a cosmological lookback time taking into account the redshift of the galaxy. The inset indicates the scale on the (log SSFR, log $t$) plane used to represent the star-formation histories in each bin. In each panel, the cyan background points represent all the  galaxies from our sample that belong to the same spectral class.

We now turn to the discussion of the star-formation histories. Concerning SF galaxies, the first thing to notice from Fig. \ref{SFH-indiv} is that, at any redshift, the SSFR  changes from slightly decreasing with time for the highest masses (log $M_\star$ $>$ 11.5) through roughly flat for intermediate masses  to increasing with time for the smaller masses. 
We also see that, for a given mass bin,  the ratio of recent to past star-formation rate increases steadily with $z$ for all mass bins. There is, of course, a certain dispersion in all those relations, as seen from the curves indicating the percentiles, but the described tendencies are unambiguous.

To interpret the observed trends, we first must ask whether, in a given mass bin, the galaxies in the panels with higher values of $z$ are the predecessors of the galaxies with lower $z$. This is not necessarily true, because of mass growth, which, in the (log $M_\star$, $z$) plane,  tends to move the galaxies towards the right as redshift decreases. This effect, however, is in general not crucial for the galaxies of our sample, given the width of our redshift bins. Indeed, for a typical SSFR of $10^{-10}$ yr$^{-1}$ as suggested by the top panel of Fig. \ref{SFH-indiv}, the mass of stars in the galaxy  doubles    during a time interval of 10 Gyr, while the width of our mass bins is 0.5 dex.  There are  episodes where the SSFR is significantly higher than $10^{-10}$ yr$^{-1}$, but, as seen in Fig.  \ref{SFH-indiv}, they are short, so they do not contribute much to the stellar mass growth.   Thus, roughly, in a given $M_\star$ bin, galaxies at higher redshifts  could be considered as the progenitors of galaxies seen at lowest redshift  (but, of course, they may change class during their evolution, for example an episodic star formation may take place or an AGN may appear). In a given mass range, aperture effects qualitatively explain the observed tendencies of the SSFR with redshift, since as redshift increases increasing portions of the star forming disk is covered by the SDSS fiber.  In addition, we must take into account the existence of the bias against red galaxies which, in a given mass bin, likely grows with increasing redshift (see Appendix \ref{app:sel}). Thus, the apparent redshift evolution of the SSFR in a given mass bin is perhaps simply due to these two causes and not an evolutionary effect.

The middle panel of Fig. \ref{SFH-indiv} is analogous to the top panel  but is for AGN hosts. Note that, because of our decision to show results only for subpanels populated by at least 50 galaxies, some panels have disappeared while a few others have appeared with respect to the SF case.  Compared to the star-formation histories of SF galaxies, those of AGN hosts show similar tendencies, with, however, the SSFRs being lower at small lookback times than those of SF galaxies in the same ($M_\star, z$) bin. We will return to this point below.

The bottom panel of Fig.  \ref{SFH-indiv} is analogous to the top panel but concerns galaxies appearing as retired instead of star-forming. Here we witness a very strong decrease of the SSFR with time in all the subpanels, with a rather small dispersion\footnote{We believe that the upturn at small lookback times seen at the smallest redshifts is an artefact due to the incomplete treatment of horizontal branch and/or blue stragglers in the BC03 evolutionary synthesis models used in the {\sc starlight} analysis. As previously shown by Ocvirk (2010) and Gonz{\'a}lez Delgado \& Cid Fernandes (2010), the lack of such old and hot phases tends to produce fake bursts in a full spectral fitting analysis.}.

\subsection{Comparison of the star-formation histories of the three galaxy spectral classes: a clue to the star formation -- AGN connection?}
\label{sfhcomp}

Galaxies that are now appearing as retired must have been seen as star-forming in the past, perhaps not continuously, but at least episodically. Thus their progenitors -- or at least some of them -- are actually in the top  (or middle) panel of Fig. \ref{SFH-indiv}. It is therefore interesting to compare in more detail the star-formation histories of our three classes of objects. 

\begin{figure}
  \centering
  \includegraphics[scale=0.4, trim={0 0 0 0mm}, clip]{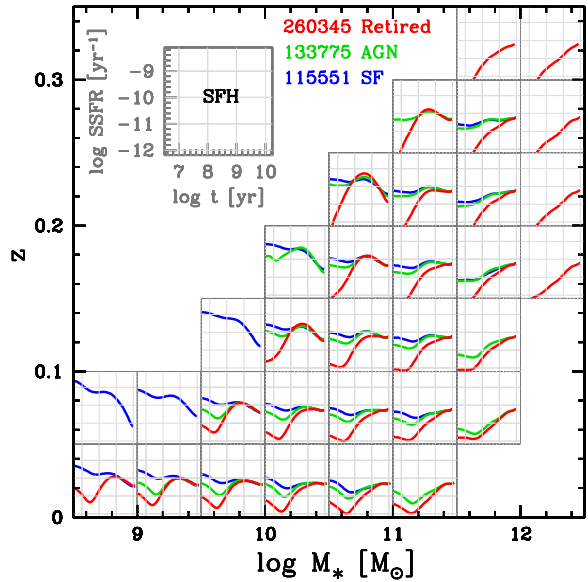}
  \caption{Star-formation histories in mass-redshift bins for all the galaxies. They are represented in blue for SF galaxies, in green for AGN hosts, and in red for retired galaxies.}
  \label{SFH-all}
\end{figure}

Fig.~\ref{SFH-all} displays the median star-formation histories of SF galaxies (blue curves), AGN hosts (green) and retired galaxies (red) in ($M_\star, z$) bins. 
One can see that,  at any redshift, the star-formation histories as obtained by {\sc starlight} for SF galaxies and AGN hosts are roughly similar. The largest differences occur at small redshift and small masses. Another thing to note is that the SSFRs of retired galaxies are identical to those of SF and AGN galaxies at the largest lookback times diverging at lookback times of one to a few Gyr.

The curves in Fig.~\ref{SFH-all} can be read as follows.  Judging from the point where the green and blue curves split apart in the figure, 
the nuclear activity kick-off  in SF galaxies reduces the level of star formation during the next 0.1--1 Gyr with respect to SF galaxies in the same ($M_\star, z$) bin. In low mass galaxies,  $M_\star \sim 10^{9-10.5} M_\odot$, once the AGN is turned on it inhibits star formation for the next $\sim$ 1 Gyr. In high mass galaxies this delay seems shorter: $\sim$ 0.1 Gyr for  $M_\star \sim 10^{11} M_\odot$\ and perhaps even less for higher masses, as can be read from Fig.~\ref{close-up} (top), which shows a close-up of Fig.~\ref{SFH-all}. All these AGNs keep forming stars, albeit at a reduced rate.

\begin{figure}
\centering
\includegraphics[scale=0.4, trim={0 0 0 0mm}, clip]{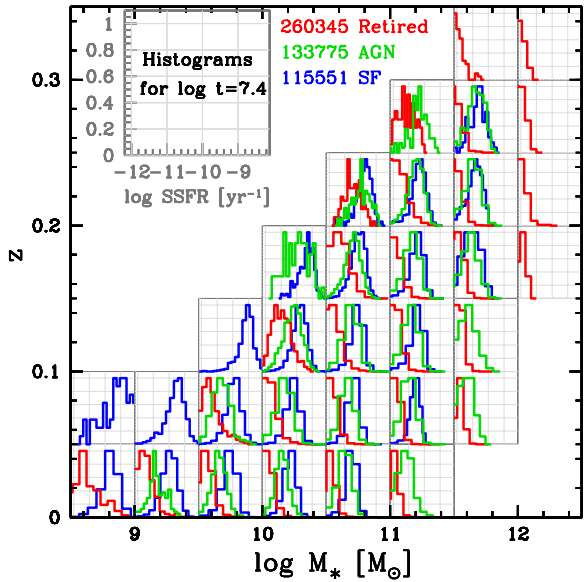}
\caption{ Distribution of recent star formation rates (at lookback time of 25 Myr) in mass-redshift bins for all galaxies. Histograms for SF galaxies are in blue, for AGN hosts in green, and for retired galaxies in red. }
 \label{SFH-hist}
\end{figure}

Another interpretation of  Fig. \ref{SFH-all} could be as follows. The population of AGN hosts could actually be composed of retired galaxies and of massive SF galaxies, resulting in  median star-formation histories of AGN hosts intermediate between SF and retired galaxies. In such a circumstance Fig.~\ref{SFH-all} would not necessarily imply the existence of a causal link between the AGN phenomenon and recent star formation.  One would then expect a dichotomy in the distribution of the star-formation rates of AGN hosts. Fig.~\ref{SFH-hist} shows the distribution of the star-formation rates at $t = 24.5$ Myr (which Asari et al. 2007 found that best matches the \Ha-calibrated SFR) for the three galaxy types. The distribution for AGN hosts is not bimodal, which rules down this second interpretation.

The overall quenching factor (as guessed from the SSFR(AGN)/SSFR(SF) ratio in a given bin) decreases with increasing $M_\star$. At low $M_\star$, AGN hosts have specific star-formation rates nearly 10 times lower than SF galaxies, while at high $M_\star$ this difference reduces to a factor of 2 or less.

 \begin{figure}
\centering
\includegraphics[scale=0.4, trim={0 0 0 0mm}, clip]{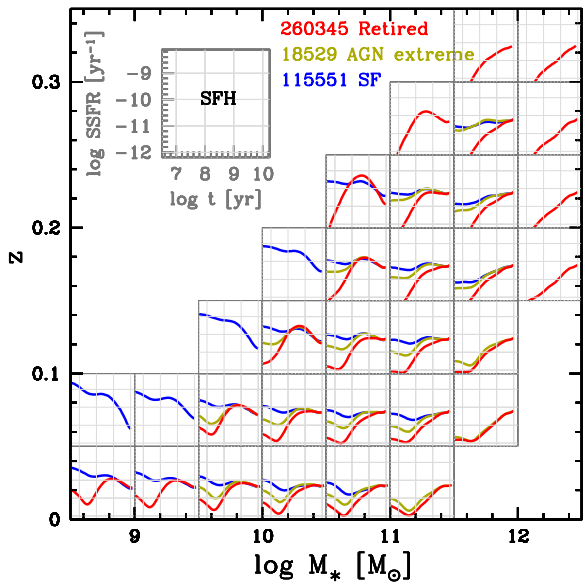}
\caption{Similar to Fig.  \ref{SFH-all} but here the AGN class contains \emph{only } galaxies which have EW(\Ha) $ > 3$\AA\  \emph{and} are above the Kewley (2001) line  in the BPT diagram.  They are represented with the gold curve.}
 \label{SFH-Kew3}
\end{figure}

 \begin{figure}
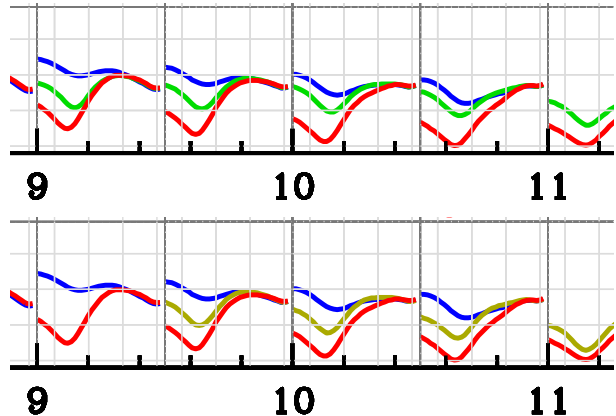

\centering
\includegraphics[scale=0.8, trim={110 32 180 156mm}, clip]{figs-reduced/SFH-all-grazy.ps}
\includegraphics[scale=0.8, trim={110 32 180 156mm}, clip]{figs-reduced/SFH-all-AGNKew3.ps}
\caption{A close-up of the lower part of Fig. \ref{SFH-all} \emph{(top)} and Fig. \ref{SFH-Kew3} \emph{(bottom)}.}
 \label{close-up}
\end{figure}

Fig.~\ref{SFH-Kew3} is analogous to Fig.  \ref{SFH-all} but here we only consider AGN hosts that are dominated by the AGN (i.e. galaxies which have EW(\Ha) $ > 3$\AA\ \emph{and} are above the K01 line in the BPT diagram.) Qualitatively, the behaviour is the same as in Fig. \ref{SFH-all}, except that the quenching factor is larger (compare top and  bottom  of Fig. \ref{close-up}).

\section{Summary}
\label{summary}

In this paper, we have addressed anew the topic of the star formation -- AGN\footnote{Note that our working definition of an AGN host is that the  galaxy contains a \emph{detectable} AGN, but the AGN itself may be weak, in particular with respect to the ionization due to massive stars in the vicinity of the nucleus. Comparison with other studies must keep this subtlety in mind.}  connection in the context of global galaxy evolution by studying several aspects. 
\begin{itemize}
  \item We have paid special attention to \emph{retired} galaxies, these AGN impostors which plague the AGN literature while they are merely galaxies that stopped forming stars, commonly mistaken for LINERs even in recent studies.
  \item We have used a fine mesh of stellar-mass bins to compare the properties of star-forming, AGN hosts and retired galaxies,  since it is known that the stellar mass is a determinant parameter for galaxy evolution. Studies that divide galaxies only with respect to their spectral types are prone to misinterpretations due to the fact that the mass distributions of the three spectral classes considered are different.
  \item We also used a fine mesh of redshift bins, which allows us to look into evolutionary and aperture effects -- although it remains difficult  to disentangle them. 
\end{itemize}

We have thoroughly investigated  the colour biases in the SDSS-DR7 Main Galaxy Sample (Appendix \ref{app:sel}): Due to its $m_r$-limited nature, the sample discriminates against \textit{ red} galaxies for decreasing galaxy masses and increasing redshifts. At the smallest reshifts, the Main Galaxy Sample misses red galaxies with masses below $10^{10} M_\odot$\, in proportions that are not possible to evaluate.

We have shown that the census of emission-line spectral types  conveys very different interpretations of galaxy evolution when using the BPT diagram or when using the WHAN. The WHAN diagram has two important advantages over the BPT:  \emph{i)} it does not set aside a large  number of the galaxies since the requirements for a galaxy to be plotted in this diagram are much smaller than for the BPT; \emph{ ii)} it does not attribute to nuclear activity the emission-line ratios seen in weak-line objects powered by HOLMES. In the remaining of our study, we therefore only considered the three classes of galaxies as defined by the WHAN diagram: pure star-forming (SF), containing a detectable AGN, and retired.

In the mass range where the Main Galaxy Sample is free of colour bias, we find that in each mass bin the proportion of SF galaxies decreases with decreasing redshift while that of retired galaxies increases. 
Such a trend is expected in  evolutionary terms (like people, galaxies evolve to retirement), but this in not an unambiguous interpretation given that aperture effects work in the same sense. 
The vast majority of galaxies with $M_\star > 10^{11.5} M_\odot$\ have formed all their stars at redshift larger than 0.4. 

We find that the population of AGN hosts is never dominant for galaxy masses larger than $10^{10} M_\odot$. Note that we reach  such a conclusion in spite of our definition of AGN host being  considerably less restrictive than the one generally used, since we consider as AGN hosts all the galaxies that contain a \emph{detectable} AGN, and not only galaxies whose ionization is \emph{dominated} by an AGN.

From the census of emission-line spectral types we have been able to infer the AGN lifetimes without any assumption on the physics of accretion on the massive black hole  nor on the radiation properties of the AGN.   For \emph{detectable} AGNs we find a lifetime  of $\lesssim  1-5$ Gyr for galaxies with masses between $10^{10}$ and  $10^{12} M_\odot$, and $\lesssim 3\times 10^8$ yr for masses above $10^{12} M_\odot$.  The lifetimes of the AGN-\emph{dominated} phases (i.e. corresponding to galaxies which have EW(\Ha) $ > 3$\AA\ \emph{and} are above the K01 line) are shorter:  a few $10^8$ yr. Of course, the lifetimes obtained in such a way  are dependent on the criterion used to define the activity -- and should be used only in similar contexts.

A deeper and independent insight into the manifold of galaxies is provided by the studies of their star-formation histories, as obtained by the spectral synthesis code {\sc starlight}.  In a star-forming galaxy, the onset of nuclear activity reduces the star-formation rate. Once the AGN is turned on it inhibits star formation for the next $\sim$ 0.1 Gyr in galaxies with  $M_\star \sim 10^{10} M_\odot$, and $\sim$ 1 Gyr in galaxies with  $M_\star \sim 10^{11} M_\odot$.  The overall quenching factor is lower for higher values of $M_\star$. In AGN-dominated galaxies, as opposed to galaxies with a \emph{detectable} AGN, the AGN phases are shorter: a few $10^8$ yr. At low $M_\star$, the quenching factor is larger.  Note that the values for the AGN lifetimes obtained from demographic arguments and from star-formation history considerations agree very well with each other.

Of course, none of this tells us anything about the physical conditions which lead a galaxy to ignite its supermassive black hole, nor how exactly this inhibits star formation. Internal (bars and other angular momentum related physics) and/or external (interactions and environment) factors should play a role.

\section*{Acknowledgments}
We thank the referee for an insightful report which helped to  improve our paper.
G.\ S., N.\ V.\ A.\ and R.\ C.\ F.\ acknowledge the support from the CAPES--COFECUB project 30007ZD. 
L.\ S.\ acknowledges the support from CAPES and FAPESP.
M.\ V.\ C.\ D.\ thanks FAPESP and CAPES/PDSE--BEX 8509/11-5 scholarships that allowed him to develop this
project. 
He also  acknowledges the hospitality of the LUTH, at Observatoire de Paris for a long term visit. 
N.\ V.\ A.\ ackowledges the support and hospitality of the LUTH, at Observatoire de Paris, for short term visits. 
This work has made use of the computing facilities of the Laboratory of Astroinformatics (IAG/USP, NAT/Unicsul), whose purchase was made possible by the Brazilian agency FAPESP (grant 2009/54006-4) and the INCT-A.

All the authors wish to thank the team of the
Sloan Digital Sky Survey (SDSS) for their dedication to a project
which has made the present work possible. Funding for the SDSS and SDSS-II has been provided by the Alfred P. Sloan Foundation, the Participating Institutions, the National Science Foundation, the U.S. Department of Energy, the National Aeronautics and Space Administration, the Japanese Monbukagakusho, the Max Planck Society, and the Higher Education Funding Council for England. The SDSS Web Site is http://www.sdss.org/.
The SDSS is managed by the Astrophysical Research Consortium for the Participating Institutions. The Participating Institutions are the American Museum of Natural History, Astrophysical Institute Potsdam, University of Basel, University of Cambridge, Case Western Reserve University, University of Chicago, Drexel University, Fermilab, the Institute for Advanced Study, the Japan Participation Group, Johns Hopkins University, the Joint Institute for Nuclear Astrophysics, the Kavli Institute for Particle Astrophysics and Cosmology, the Korean Scientist Group, the Chinese Academy of Sciences (LAMOST), Los Alamos National Laboratory, the Max-Planck-Institute for Astronomy (MPIA), the Max-Planck-Institute for Astrophysics (MPA), New Mexico State University, Ohio State University, University of Pittsburgh, University of Portsmouth, Princeton University, the United States Naval Observatory, and the University of Washington.

\appendix

\section{About selection effects in the SDSS Main Galaxy Sample}
\label{app:sel}

\begin{figure}
\centering
\includegraphics[scale=0.1, trim={0 0 0 0mm}, clip]{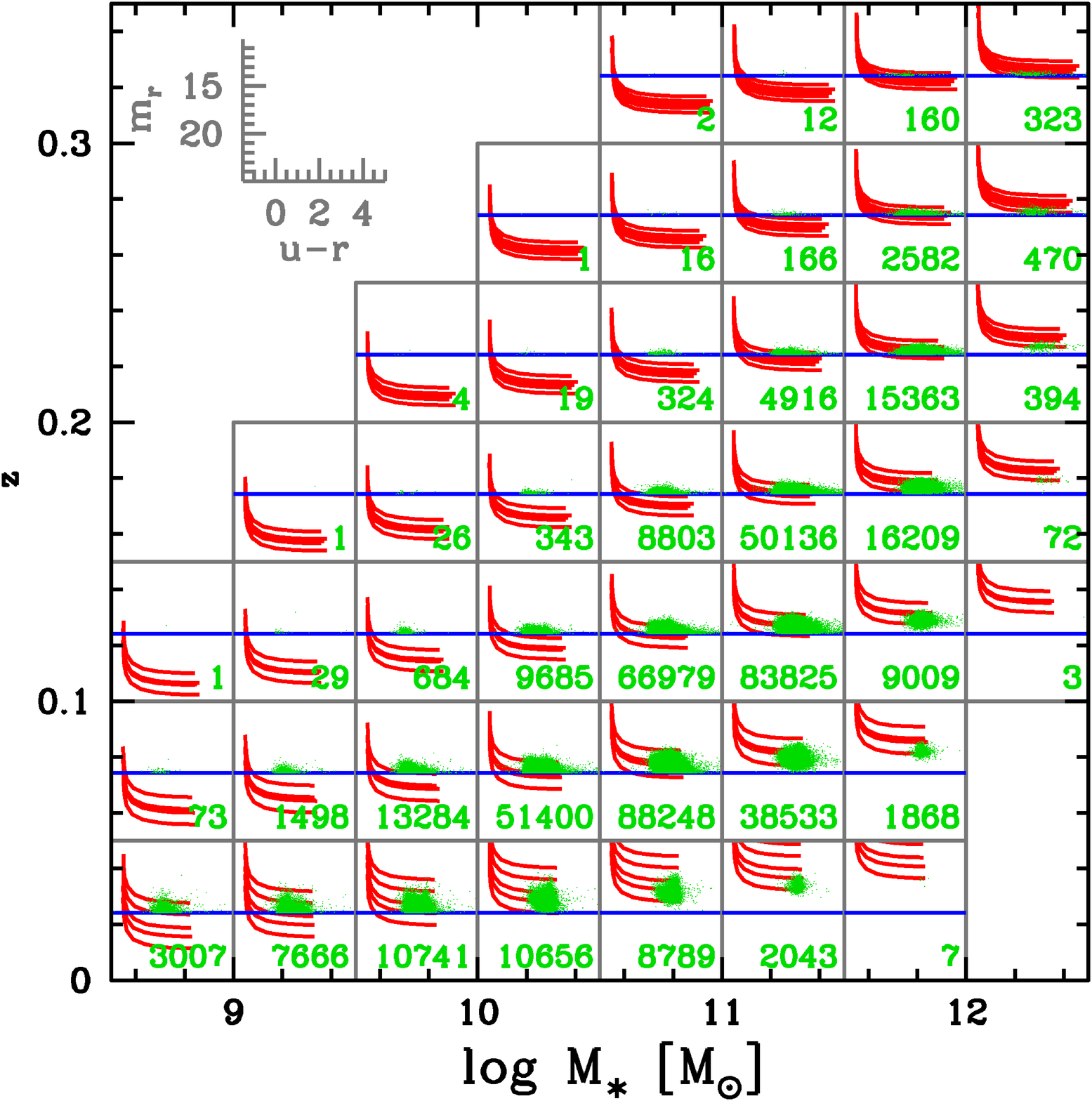}
\includegraphics[scale=0.1, trim={0 0 0 0mm}, clip]{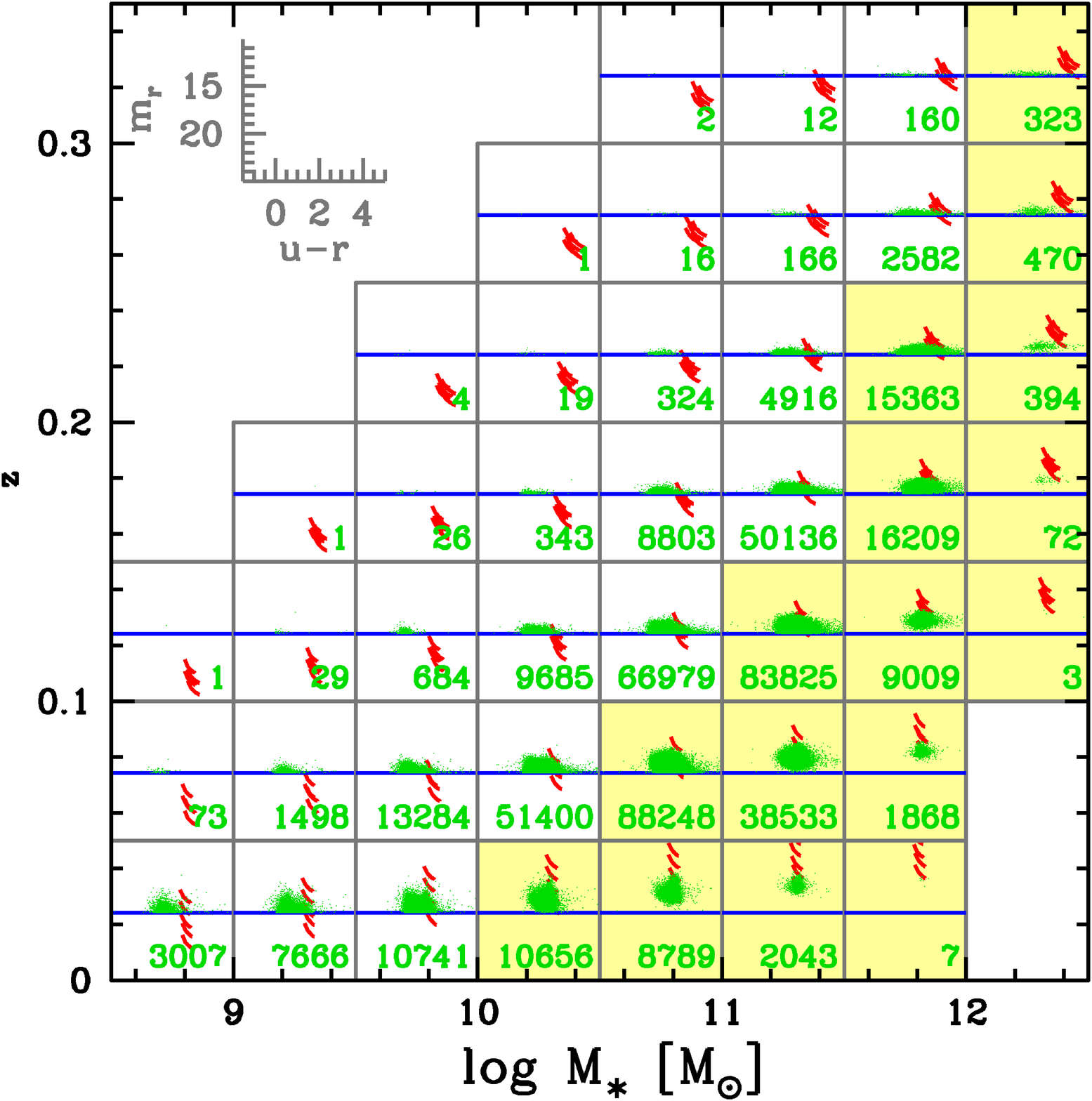}
\caption{Colour-magnitude diagrams of galaxy toy-models in bins of mass and redshift. \emph{Top:} models M1; \emph{bottom:} models M2. The inset indicates the magnitude scales. The blue horizontal line indicates the limiting magnitude $m_r = 17.77$ of the Main Galaxy Sample. In each subpanel, the corresponding number of WHAN sample galaxies is indicated and  their colour-magnitude diagram is plotted in green.  }  
 \label{sel_WHAN}
\end{figure}

Here we analyze whether in our sample there is a selection effect for or against certain galaxy colours due to the fact that the Main Galaxy Sample (MGS) of the SDSS is limited in $r$ apparent magnitude to $m_r > 17.77$\footnote{The Main Galaxy Sample also imposes a cut on the half-light surface brightness ($\mu_{50}$) of 24.5 mag arcsec$^{-2}$. After applying the $m_r$ criterion, only 1\% of the objects classified as galaxies in the SDSS fall into the low surface brightness region ($23.0 < \mu_{50} < 24.5$; Strauss et al. 2002), so this has virtually no impact on our study.}. For that, in a first approximation we consider a toy-model (model M1) in which a galaxy is composed of a very young stellar population of 1 Myr and of an old stellar population of 13 Gyr and we vary the proportion in mass of those two stellar populations. The SSP models are solar metallicity models taken from Bruzual \& Charlot (2003). For each proportion of the young and old stellar populations, we compute the resulting colour $u-r$, applying the response curves of the SDSS $u$ and $r$ filters. We then compute the absolute magnitude in the $r$ band corresponding to a given value of $M_\star$, and the apparent magnitude $m_r$ corresponding to a given redshift $z$.  The top panel of Fig.  \ref{sel_WHAN} shows the colour-magnitude diagram for the models M1 inside each of our mass-redshift bins. For each subpanel, the red curves represent sequences of models for the central values of $M_\star$ and $z$ and for the formal extreme values for the considered bin, hence we have five curves inside each subpanel (the models are shown only for bins containing at least one object from our WHAN subsample). The SDSS Main Galaxy Sample limit ($m_r = 17.77$) is marked by the blue horizontal lines. In each subpanel the green points show the corresponding \emph{observed} colour-magnitude diagram (the effect of extinction correction is minimal on the overall diagram).

The first thing to notice in  Fig. \ref{sel_WHAN}(top) is that the red curves reach a plateau as soon as  $u-r >1$ mag, which effectively means that, for the colours seen in the our sample, $m_r$ is in practice a measure of a galaxy's stellar mass. We also note that in a certain number of bins, all the red curves are above the blue line, meaning that 
the sample is complete for those bins even taking into account the dust attenuation\footnote{The distribution of the line-of-sight attenuation parameter $A_V$ peaks around 0.0 and 0.5 mag for retired and star-forming galaxies  respectively, while the nebular $A_V$ for the Balmer decrement is consistently around $1.0$ mag for the emission-line retired and  varies from $1.0$ in the highest-redshift bin to $2.0$ mag in the lowest-redshift bin, as can be seen in Figs. \ref{AV_WHAN} and \ref{AV_lines_WHAN} of Appendix \ref{add}.  This means that the correction for attenuation is of the order of 1 mag, which is very small compared to the $m_r$ range in our toy models.}. This is the case for log $M_\star > 10 $ in the smallest redshift bin, for log $M_\star > 10.5 $  for $z < 0.1$, for log $M_\star > 11 $  for $z < 0.15$ etc. In addition, for those bins where the sample is not complete, i.e. the bins for which the domain of the red curves is crossed by the blue line, the slope of the red curves is zero. This means that, even in bins where the sample is not complete, there is no bias against any particular colour.  One should however be cautious with   few bins where the sample is missing a large fraction of objects, i.e. log $M_\star < 9.5 $  for $z > 0.05$, log $M_\star <10 $  for $z > 0.1$, log $M_\star <10 $  for $z > 0.15$, log $M_\star <11 $  for $z > 0.2$.

However, our toy model combining one very old (13 Gyr) and one very young (1 Myr) stellar population is not the most appropriate to represent the majority of  old galaxies. We therefore computed a second toy model (model M2) in which the SSP to be combined to the 13 Gyr one has an age of 2.5 Gyr. The results are shown in the bottom panel of Fig. \ref{sel_WHAN}. Of course, in this case, all the models are red and do not represent the bulk of the Main Galaxy Sample. But they show that the Main Galaxy Sample certainly misses red objects at low values of  $M_\star$ for the smallest redshifts, with increasing limits on  $M_\star$  for increasing redshifts. 

It is unfortunately impossible to quantify in each  ($M_\star, z$) bin the  bias against red galaxies that is present in the  Main Galaxy Sample due to the limitation in $m_r$. What we can do is to compare our toy models with the observations in Fig.  \ref{sel_WHAN}(bottom) to judge which are the ($M_\star, z$) bins which are definitely complete (these are indicated the  yellow background  in the figure\footnote{An additional potential issue is the contribution of the \Ha\ line emission to the $m_r$. However, given the colours of the galaxies in the considered sample, the effect -- as can be judged from the computations by Garc{\'{\i}}a-Vargas et al. (2013) -- is minor. In the most extreme cases, it would act in favour of the bluest and least massive galaxies.}). Adjacent bins are probably only slightly incomplete in red objects, while the bias may become non negligible as $M_\star$ decreases. For example, we could be missing a large amount of  red galaxies with $M_\star < 10^{9} M_\odot$.

\section{A warning on stacking galaxy spectra}
\label{app:stacks}

\begin{figure*}
\centering
\includegraphics[width=0.40\textwidth, trim=0 0 30 20, clip]{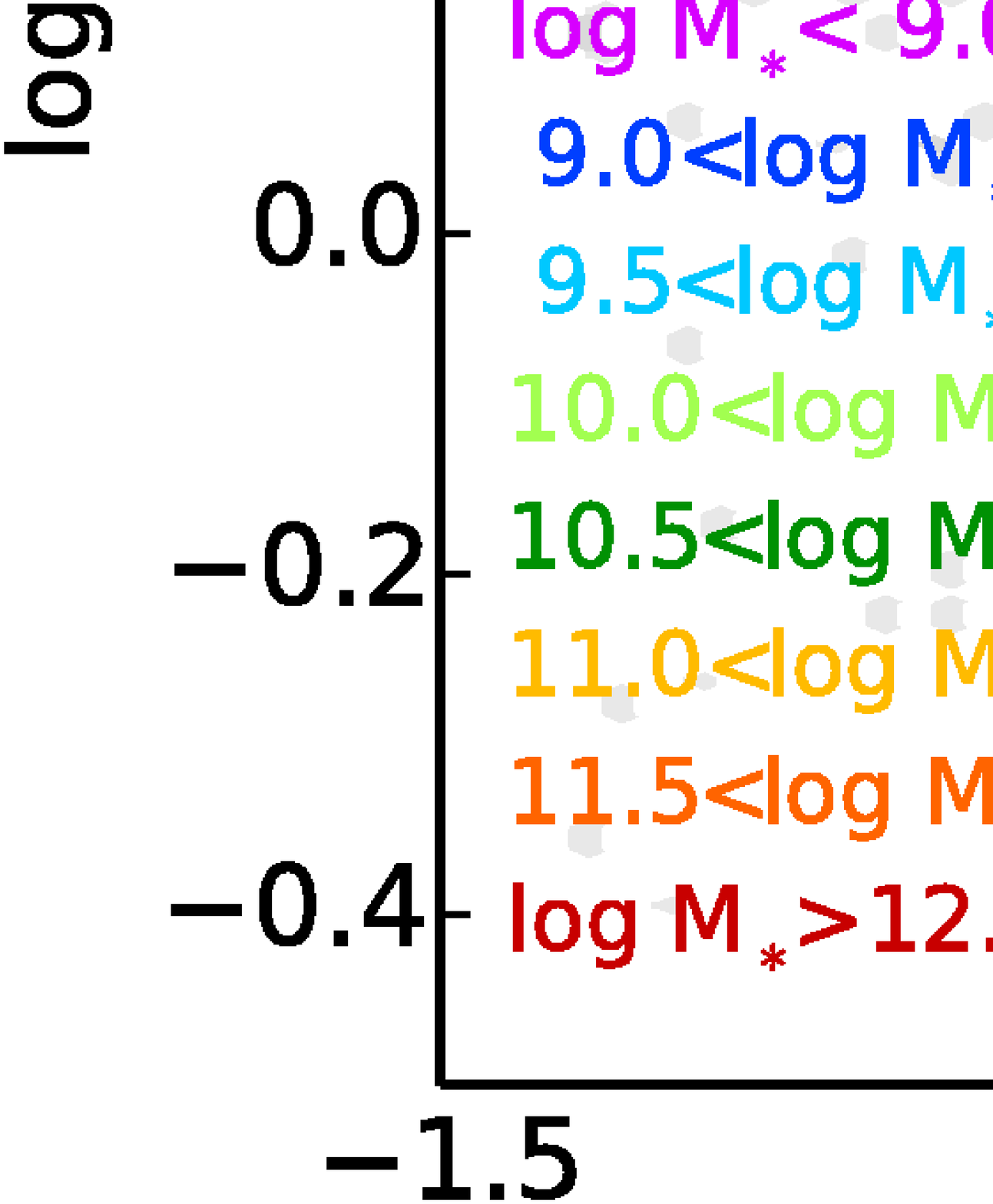}
\includegraphics[width=0.40\textwidth, trim=0 0 30 20, clip]{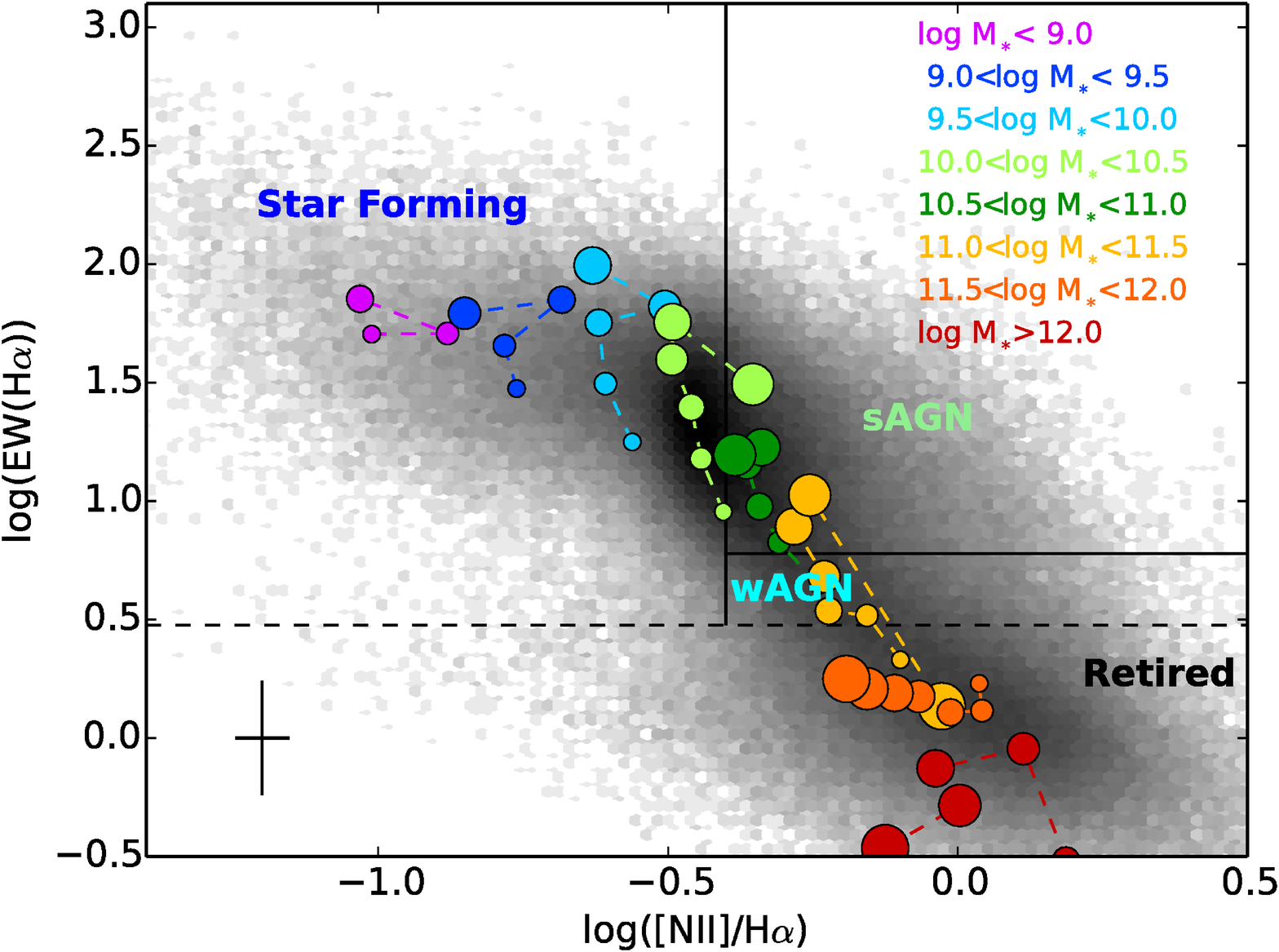}
\includegraphics[width=0.40\textwidth, trim=0 0 30 20, clip]{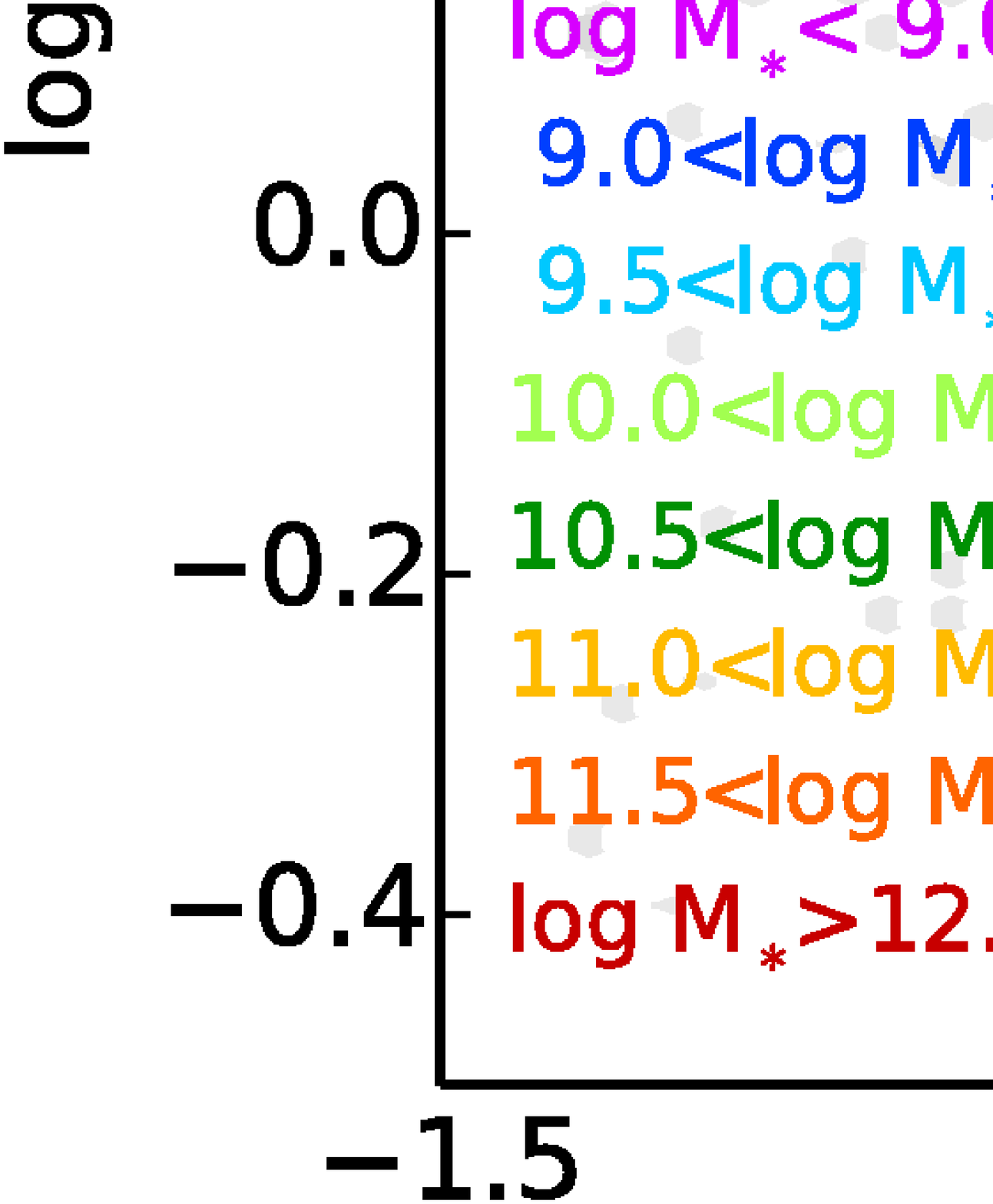}
\includegraphics[width=0.40\textwidth, trim=0 0 30 20, clip]{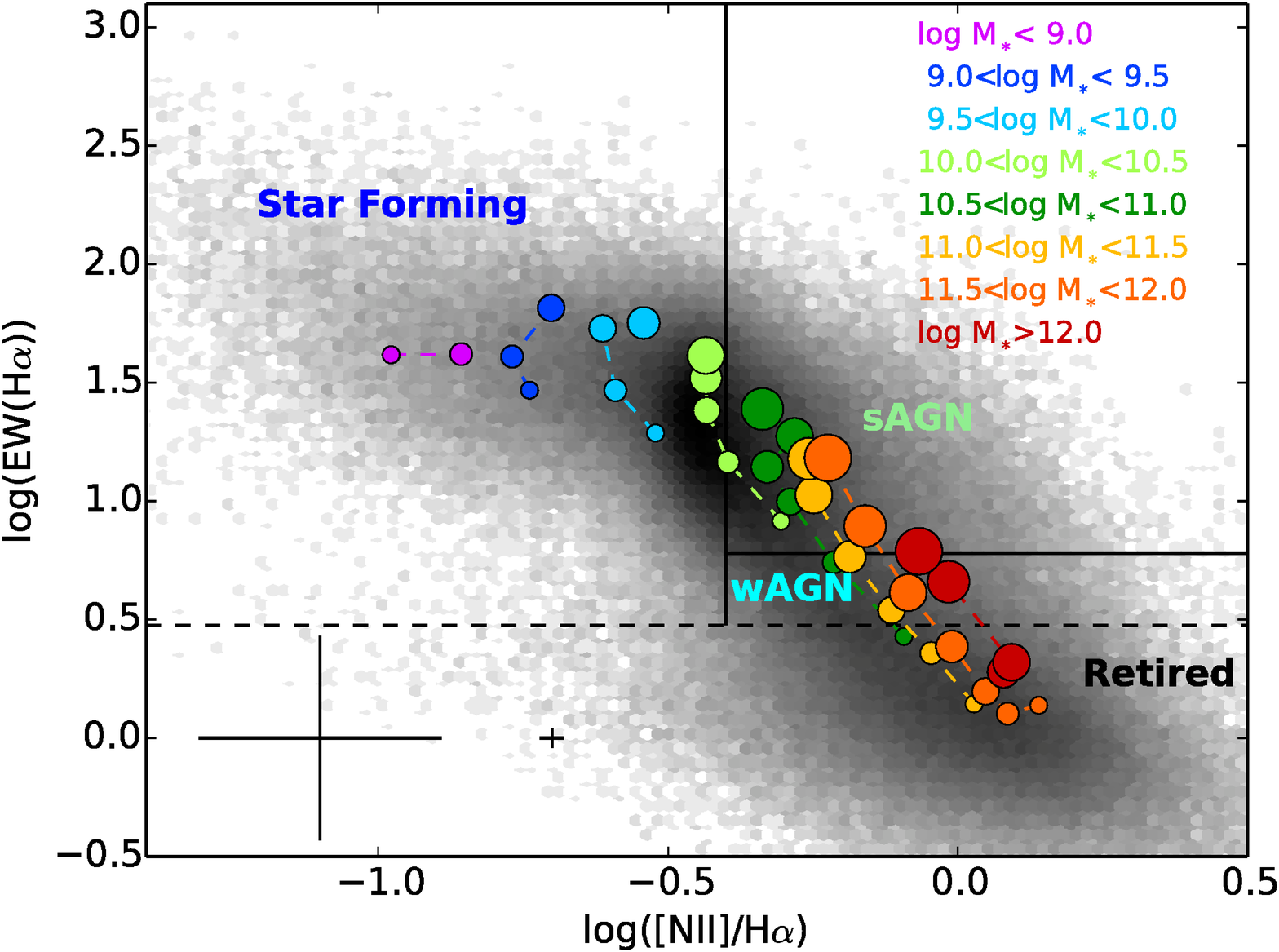}
\caption{Locations in the BPT and WHAN diagrams of galaxies with redshifts between $0.002 < z < 0.35$ grouped in mass-redshift bins. Larger circles represent higher redshifts. The different colours represent the different mass bins, and thin dashed lines join points in the same mass bin. The typical uncertainty is drawn on the top right of each panel.  The red curve on the BPT represents the S06 delimitation between pure star-forming galaxies and galaxies containing an AGN. The full and dashed black lines on the WHAN are the demarcation of different galaxies as in Cid Fernandes et al.\ (2010).  \emph{Top panels:} locations obtained from the stacked spectra. The error bars are the uncertainty in the line fluxes from the gaussian fitting code. \emph{Bottom panels:} locations obtained from the averages of individual line ratios in each bin, considering only galaxies with good spectra. The larger error bar is the standard deviation and the smaller one is the standard deviation of the mean. The background density map represents the location of the individual positions of the galaxies in our parent sample.}
\label{stacks}
\end{figure*}

In the main body of this paper we have studied the dispersions of galaxies properties within given mass-redshift bins and shown that one single bin can be populated by galaxies with completely different excitation sources. In this appendix we warn against misinterpretations from common approaches using  average galaxy properties or spectral stacking (Dobos et al. 2012; Vitale et al. 2013; Juneau et al. 2014).  What follows is a didactic exercise to show what would be concluded  had we chosen to use these approaches.

We divide our parent sample in mass and redshift bins as in the rest of the paper. The only change is to extend the extreme mass bins to include galaxies with $\log M_\star < 9$ and $\log M_\star > 12$. We then consider each mass-redshift bin as a single point to be plotted on the BPT or the WHAN diagram.

First we produce representative spectra in each mass-redshift bin by stacking all the spectra pertaining to the bin. This allows one to provide a synthetic view of the galaxy spectra in the ($M_\star, z$) space. Stacking is generally used to enhance the signal-to-noise ratio of spectra of objects believed to belong to the same category, and, for example, would be needed to obtain  accurate star-formation histories with {\sc starlight} for galaxies at higher redshifts. This is not necessary for SDSS galaxies, since individual SF histories have been obtained but it is nevertheless interesting to comment on the biases that stacking may produce.  Obviously, stacking erases information on individual galaxies and, if, in spite of our fine meshing, each bin contains galaxies of different types, the stacked spectrum may not correspond to any galaxy known. We anticipate that if, as is currently accepted, mass is the main driver of galaxy evolution, the continuum in the stacked spectrum should represent the stellar content in all the concerned galaxies reasonably well. On the other hand, the emission lines can be due to a variety of causes, most of them occuring on a short time-scale, the most obvious being current star formation or nuclear activity, and their intensities may vary significantly within a bin.

Before stacking, we normalize the spectra to the same value of  the stellar continuum at a restframe wavelength of $4010-4060$\AA.  Since even in a given redshift bin the wavelength ranges vary according to the redshift of each object, we stack the spectra considering only the wavelength range common to all the spectra in the bin.  Other ways to stack spectra would have been to use median stacking or light-weighted averages instead of average stacking. Independent of the method, the information of the population of galaxies within a bin would be erased.

The top panels of Fig. \ref{stacks} show the position in the BPT and WHAN diagrams of fictive objects created by our stacking procedure. For the BPT we show sample B and for the WHAN sample W. Each colour represents a different mass bin, following the palette used in Fig. \ref{fraction-BPT-redshift} and \ref{fraction-WHAN_all-redshift}, and the size of the symbols correspond to the redshift (size increases with increasing redshift). 
The points are rather dispersed, but some general trends are visible. From the BPT diagram one might have the impression that galaxy masses determine if a galaxy has ongoing star-formation, has an important AGN or is in between. From the WHAN diagram, we might say that the retired galaxies dominate the highest masses, the AGN hosts dominate intermediate masses and SF galaxies dominate low masses \emph{at all redshifts} studied here.

Another way to represent a collection of galaxies in an emission-line diagnostic diagram -- if the spectra are, like in the SDSS, of good enough quality to allow the measurement of emission-line intensities for each galaxy individually -- is to plot for each ($M_\star$, $z$) bin a point whose coordinates are the averages of \oiii/\Hb\, \nii/\Ha\ or EW(\Ha) for all the objects with sufficiently good quality. This way, we are selecting a different sample from the one before and giving more weight to emission-line galaxies.  The result is shown in the bottom panel of Fig. \ref{stacks}.  The points are far less dispersed, but the interpretations of the trends for ``fictive'' galaxies has not changed by much, except for the fact that  the two highest-mass bins now show the same redshift trends as the lower mass bins.

The puzzling behaviour of average properties seen above, either from stacking or from combining individual line measures, calls for great care when interpreting those results. In the ideal case, one should  examine the \textit{dispersion} of properties inside a bin, which is possible with SDSS data. For example, certain galaxies do not display emission lines at all, yet they enter in the construction of stacked spectra but not on the average emission-line properties inside a bin. In some mass bins, a fraction of the galaxies can be pure star-forming (except perhaps for an old bulge), while others may contain an AGN. It is important to study the distribution of the emission-line properties in a case where -- as here -- this is possible.

Indeed, when going to higher redshift, one might not be able to do this and will have to rely only on stacked spectra. So examining the dispersion of properties in cases where it is possible is an important way to educate oneself about possible misinterpretations of stacked data.

\section{Additional material}
\label{add}

\begin{figure}
\centering
\includegraphics[scale=0.45, trim={0  0 0 50mm}, clip]{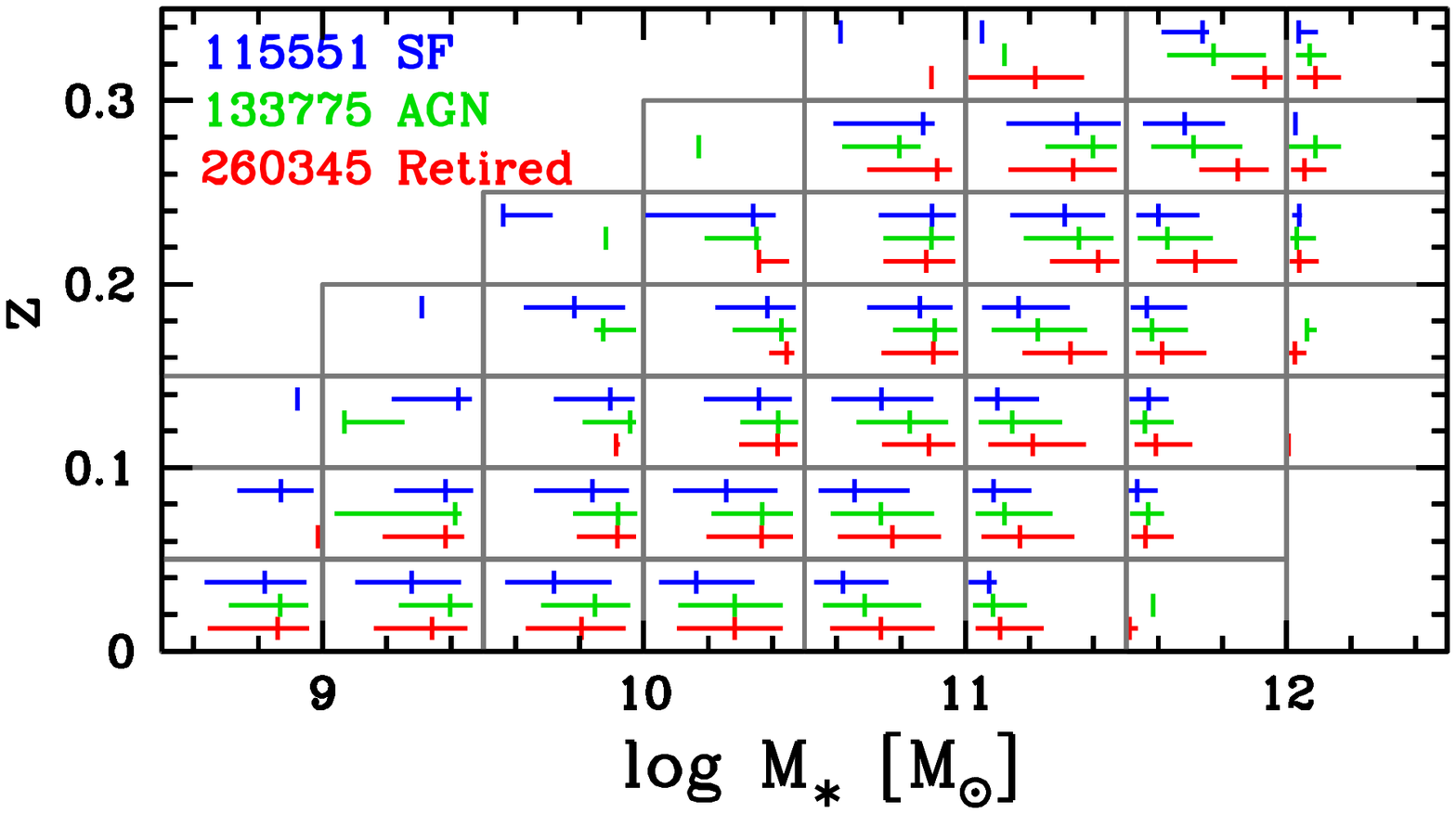}
\caption{Distribution of stellar masses in each ($M_\star$, $z$) bin. The extremities of the horizontal segments (arbitraly vertically shifted for clarity) indicates the 16th and 84th percentiles of the $M_\star$ distribution, while the vertical tickmark indicates the median value.   The different classes of galaxies are represented with the following colours. Blue: SF, green: AGN and red: retired. }
 \label{MZR_Hist_WHAN}
\end{figure}

 \begin{figure}
\centering
\includegraphics[scale=0.45, trim={0 0 0 50mm}, clip]{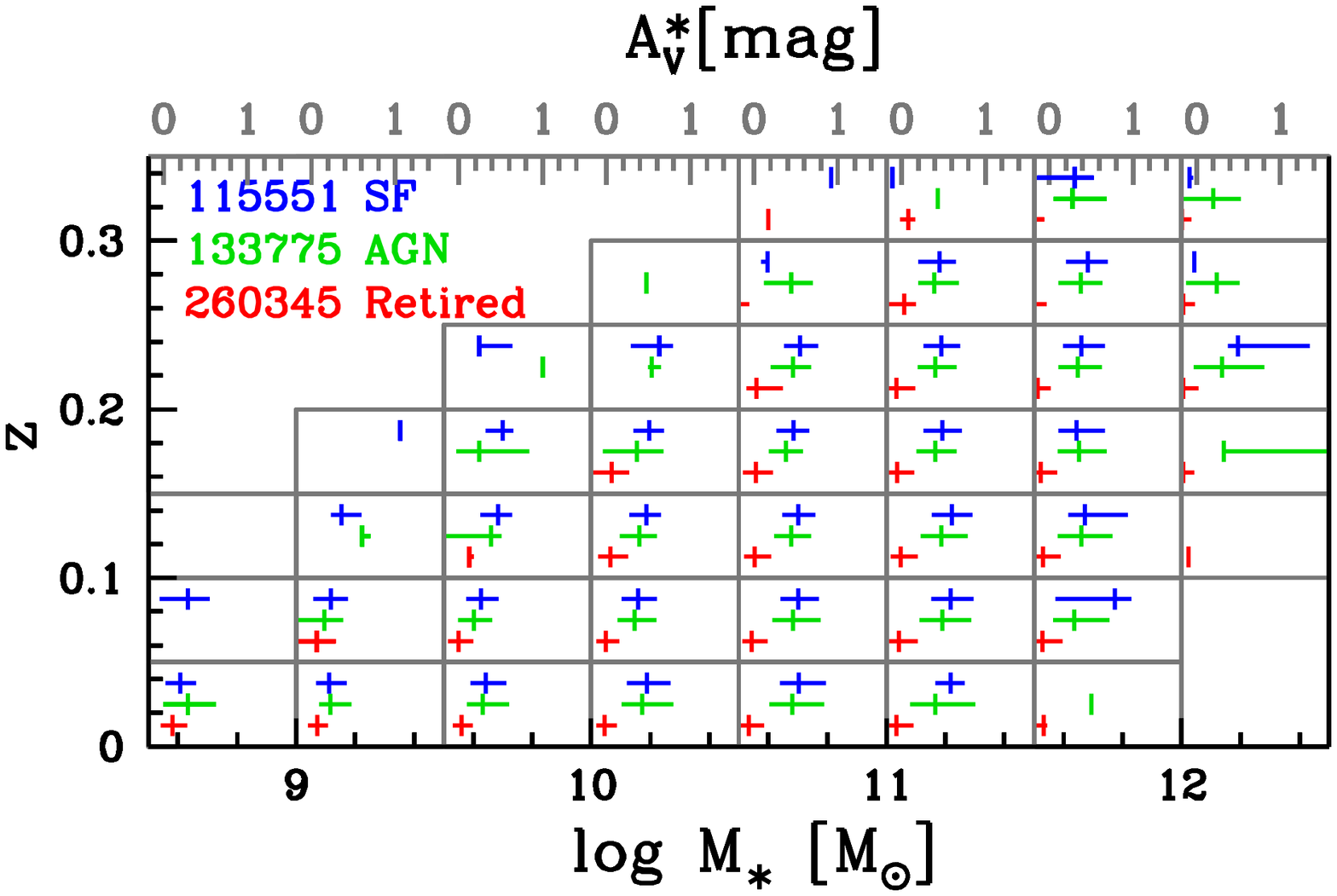}
\caption{Distribution of values of $A_V$, the extinction obtained from the {\sc starlight} analysis, in each stellar masses in each ($M_\star$, $z$) bin. The extremities of the horizontal segments indicates the 16th and 84th percentiles, while the vertical tickmark indicates the median value.  The grey-coloured scale at the top of the diagram indicates the scale for $A_V$ in each mass bin.  The different classes of galaxies are represented with the following colours. Blue: SF, green: AGN and red: retired.  }
\label{AV_WHAN}
\end{figure}

 \begin{figure}
\centering
\includegraphics[scale=0.45, trim={0 0 0 50mm}, clip]{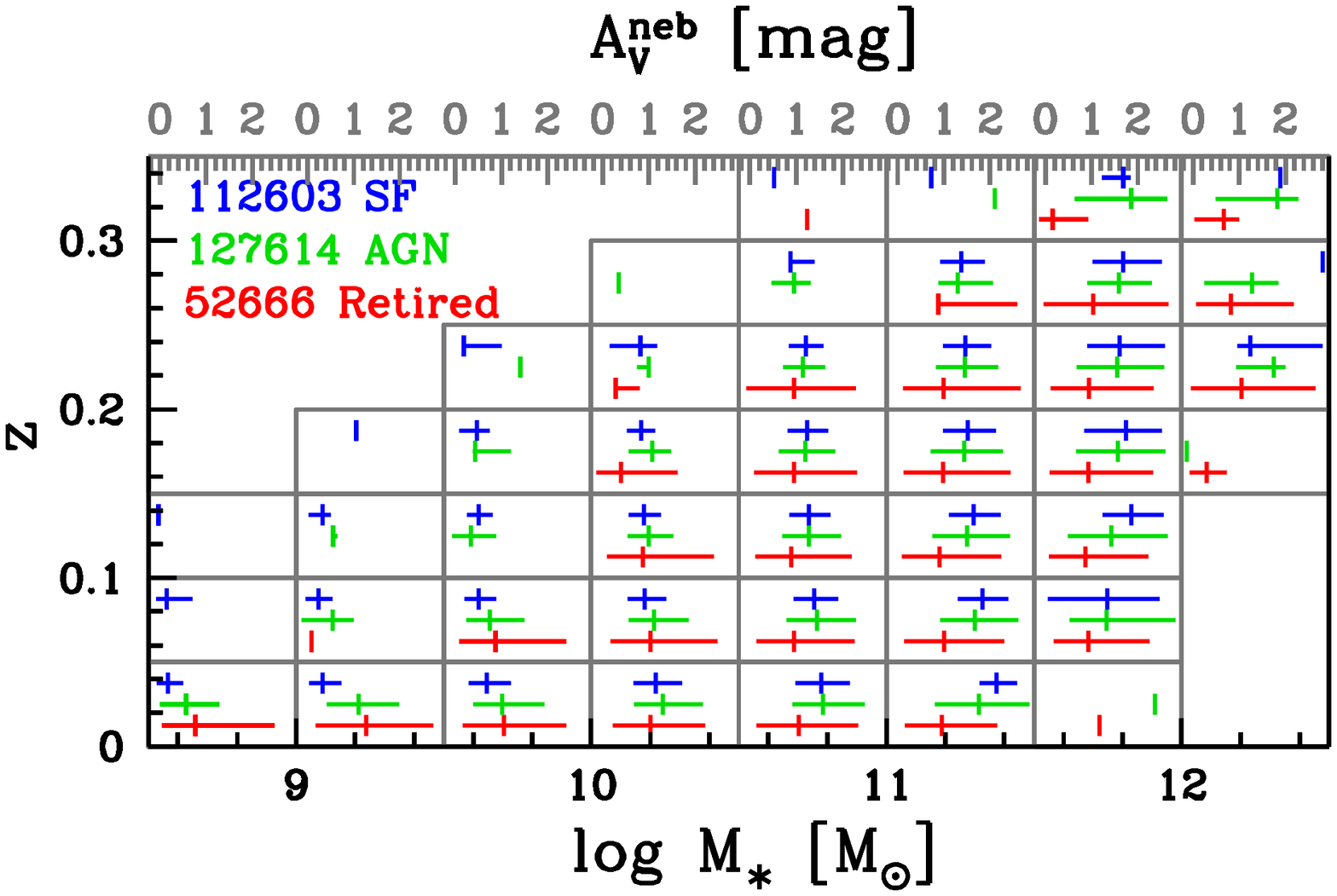}
\caption{Same as Fig.  \ref{AV_WHAN} for the extinction  $A_V$ derived from the Balmer decrement, when both \Ha\ and \Hb\ are observed with a  signal-to-noise ratio larger than 3. }
\label{AV_lines_WHAN}
\end{figure}

 \begin{figure}
\centering
\includegraphics[scale=0.45, trim={0 0 0 50mm}, clip]{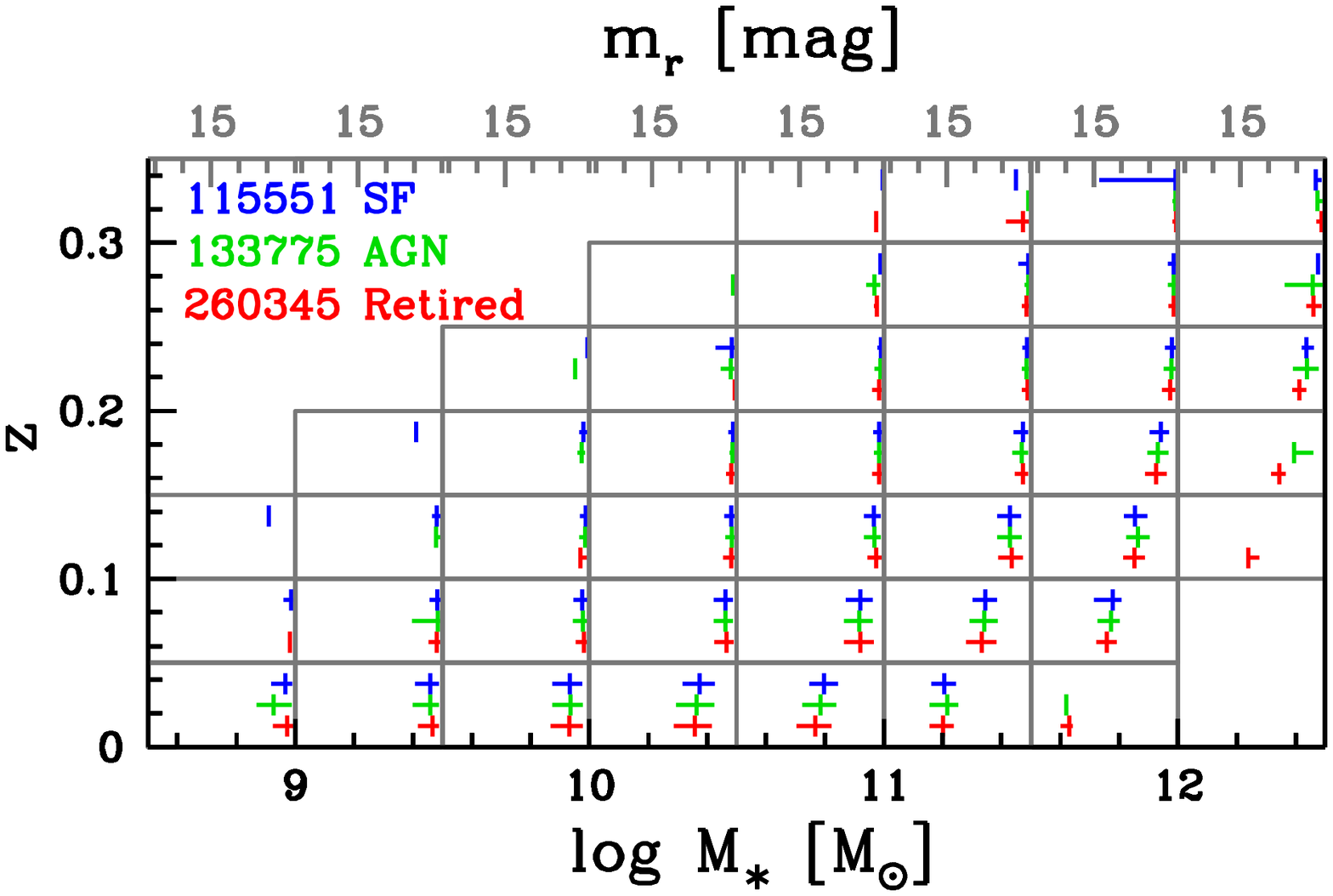}
\caption{Distributions of  $m_r$ for the various galaxy spectral types.}
\label{m_r_1_WHAN}
\end{figure}

\begin{figure}
\centering
\includegraphics[scale=0.45, trim={0 0 0 50mm}, clip]{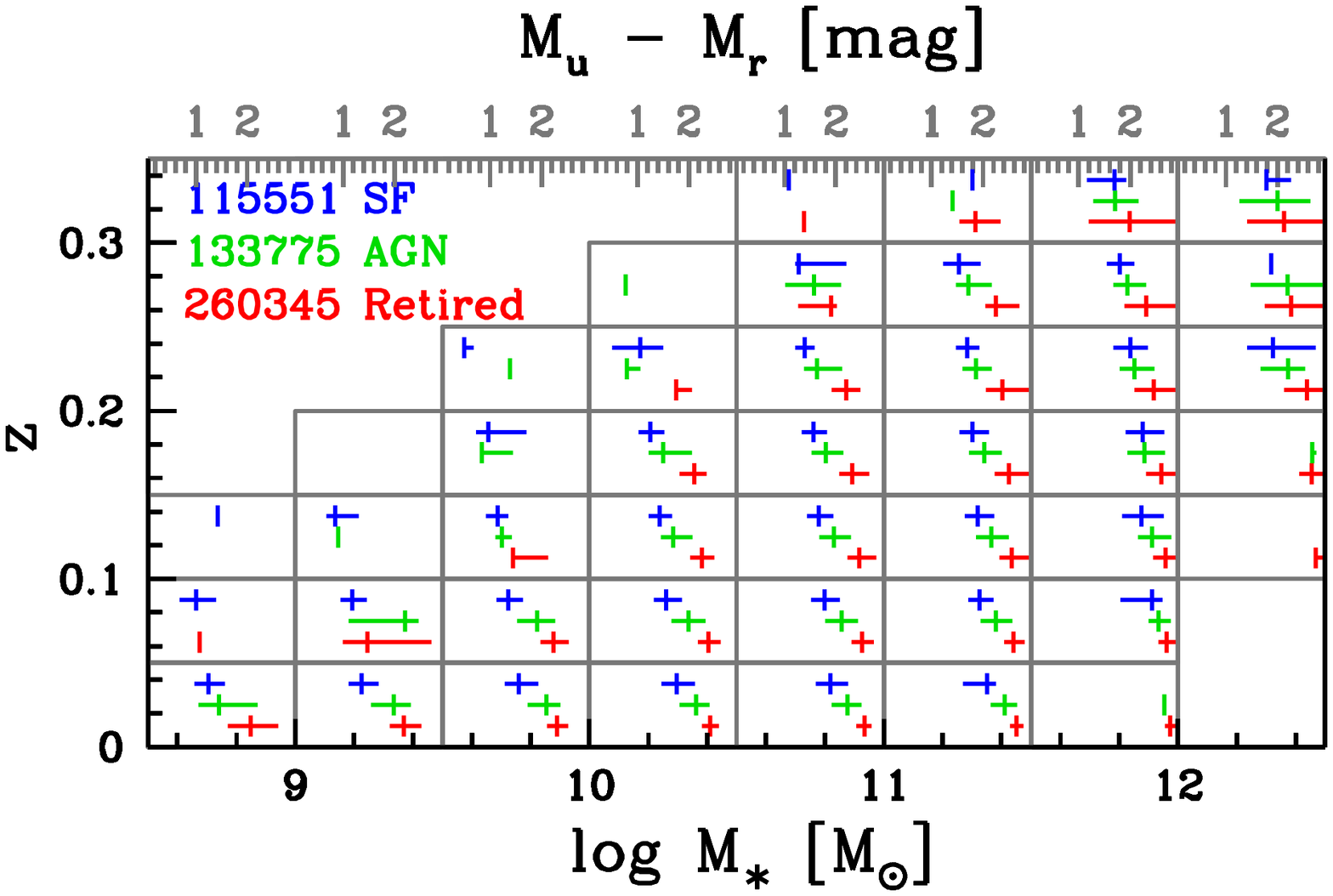}
\caption{Distributions of  the $u-r$ colours for the various galaxy spectral types.}
\label{MuMr_WHAN}
\end{figure}

To aid the  discussion of the star-formation histories in mass and redshift bins, it is instructive to visualize the distributions of certain properties of the galaxy classes considered, in every ($M_\star, z$) bin. All the figures below follow the same colour scheme: blue for SF galaxies, green for AGN hosts and red for retired galaxies. The considered sample is sample W.

Even  in the case of our very fine $M_\star$ and $z$ gridding, differences may appear in the $M_\star$ distributions  of the different classes of galaxies considered, and, for a same class, for different values of $z$. 
 Figure \ref{MZR_Hist_WHAN} summarizes the distribution of stellar masses in each ($M_\star$, $z$) bin. The extremities of the horizontal segments indicates the 16th and 84th percentiles of the $M_\star$ distribution, while the vertical tickmark indicates the median value.    We see that, although the mass distributions are not identical in each subpanel, the median values for the SF, AGN and retired classes are very close to each other.

Figures \ref{AV_WHAN} and \ref{AV_lines_WHAN} show the distribution of the stellar extinction obtained from the {\sc starlight} analysis, and the Balmer-line extinction, respectively. The fact that the the distributions of  Balmer-line extinctions are broad for the retired galaxies is actually due to the large uncertainties of this parameter for this category of objects.

Figure \ref{m_r_1_WHAN} shows the distributions of  $m_r$ for the various galaxy spectral types. It illustrates the increasing effect of the observational cut-off with increasing redshift. 

Figure  \ref{MuMr_WHAN} shows the distributions of the galaxy $u-r$ colours. We see that in each mass-redshift bin, the $u-r$ values of retired galaxies exceed those of AGN hosts by typically 0.2 dex, and that those of the latter exceed those of SF galaxies also by about 0.2 dex.

\label{lastpage}


\begin{thebibliography}{99}

\bibitem[Abazajian et al.(2009)]{Abazajianetal2009}
{Abazajian}, K.~N., {Adelman-McCarthy}, J.~K., {Ag{\"u}eros}, M.~A. et
al., 2009, \apjs, 182, 543

\bibitem[\protect\citeauthoryear{Asari et al.}{2007}]{2007MNRAS.381..263A} 
Asari N.~V., Cid Fernandes R., Stasi{\'n}ska G., Torres-Papaqui J.~P., 
Mateus A., Sodr{\'e} L., Schoenell W., Gomes J.~M., 2007, MNRAS, 381, 263 



\bibitem[Baldwin, Phillips and Terlevich(1981)]{BPT81}
{Baldwin}, J.~A. and {Phillips}, M.~M. and {Terlevich}, R.,1981, \pasp, 93, 5



\bibitem[\protect\citeauthoryear{Bamford et 
al.}{2009}]{2009MNRAS.393.1324B} Bamford S.~P., et al., 2009, MNRAS, 393, 
1324 


\bibitem[\protect\citeauthoryear{Bertelli et 
al.}{1994}]{1994A&AS..106..275B} Bertelli G., Bressan A., Chiosi C., Fagotto F., Nasi E., 1994, A\&AS, 106, 275



\bibitem[Bruzual \& Charlot(2003)]{BC03}
{Bruzual}, G. and {Charlot}, S., 2003, \mnras, 344, 1000

\bibitem[\protect\citeauthoryear{Brinchmann et 
al.}{2004}]{2004MNRAS.351.1151B} Brinchmann J., Charlot S., White S.~D.~M., 
Tremonti C., Kauffmann G., Heckman T., Brinkmann J., 2004, MNRAS, 351, 1151 

\bibitem[Cardelli et al.(1989)]{Cardellietal1989}
{Cardelli}, J.~A., {Clayton}, G.~C., {Mathis}, J.~S., 1989, \apj, 345, 245

\bibitem[\protect\citeauthoryear{Chabrier}{2003}]{2003PASP..115..763C} 
Chabrier G., 2003, PASP, 115, 763 

\bibitem[Cid Fernandes et al.(2005)]{CidFernandesetal2005}
{Cid Fernandes}, R., {Mateus}, A., {Sodr{\'e}}, L. et al., 2005, \mnras, 358, 363

\bibitem[\protect\citeauthoryear{Cid Fernandes et 
al.}{2007}]{2007MNRAS.375L..16C} Cid Fernandes R., Asari N.~V., Sodr{\'e} 
L., Stasi{\'n}ska G., Mateus A., Torres-Papaqui J.~P., Schoenell W., 2007, 
MNRAS, 375, L16


\bibitem[\protect\citeauthoryear{Cid Fernandes et 
al.}{2010}]{2010MNRAS.403.1036C} Cid Fernandes R., Stasi{\'n}ska G., 
Schlickmann M.~S., Mateus A., Vale Asari N., Schoenell W., Sodr{\'e} L., 
2010, MNRAS, 403, 1036 




\bibitem[Cid Fernandes et al.(2011)]{CidFernandesetal2011}
{Cid Fernandes}, R. and {Stasi{\'n}ska}, G. and {Mateus}, A. et al., 2011, \mnras, 413, 1687C

\bibitem[\protect\citeauthoryear{Constantin et 
al.}{2009}]{2009ApJ...705.1336C} Constantin A., Green P., Aldcroft T., Kim 
D.-W., Haggard D., Barkhouse W., Anderson S.~F., 2009, ApJ, 705, 1336 

\bibitem[Cowe et al.(1996)]{Coweetal1996}
{Cowie}, L.~L. and {Songaila}, A. and {Hu}, E.~M. et al., 1996, \aj, 112, 839


\bibitem[\protect\citeauthoryear{Deng et al.}{2012}]{2012PASJ...64...93D} 
Deng X.-F., Wu P., Qian X.-X., Luo C.-H., 2012, PASJ, 64, 93 


\bibitem[\protect\citeauthoryear{Dobos et al.}{2012}]{2012MNRAS.420.1217D} 
Dobos L., Csabai I., Yip C.-W., Budav{\'a}ri T., Wild V., Szalay A.~S., 
2012, MNRAS, 420, 1217 

\bibitem[Ekholm et 
al.(2001)]{2001A&A...368L..17E} Ekholm, T., Baryshev, Y., Teerikorpi, P., Hanski, M.~O., \& Paturel, G.\ 2001, \aap, 368, L17 

\bibitem[\protect\citeauthoryear{Fabian}{2012}]{2012ARA&A..50..455F} Fabian A.~C., 2012, ARA\&A, 50, 455 

\bibitem[\protect\citeauthoryear{Garc{\'{\i}}a-Vargas, Moll{\'a}, 
\& Mart{\'{\i}}n-Manj{\'o}n}{2013}]{2013MNRAS.432.2746G} Garc{\'{\i}}a-Vargas M.~L., Moll{\'a} M., Mart{\'{\i}}n-Manj{\'o}n M.~L., 2013, MNRAS, 432, 2746 

\bibitem[\protect\citeauthoryear{Gerssen, Wilman, 
\& Christensen}{2012}]{2012MNRAS.420..197G} Gerssen J., Wilman D.~J., Christensen L., 2012, MNRAS, 420, 197 


\bibitem[\protect\citeauthoryear{Gonz{\'a}lez Delgado et 
al.}{2009}]{2009Ap&SS.320...61G} Gonz{\'a}lez Delgado R.~M., Mu{\~n}oz Mar{\'{\i}}n V.~M., P{\'e}rez E., Schmitt H.~R., Cid Fernandes R., 2009, Ap\&SS, 320, 61 

\bibitem[\protect\citeauthoryear{Gonz{\'a}lez Delgado 
\& Cid Fernandes}{2010}]{2010MNRAS.403..797G} Gonz{\'a}lez Delgado R.~M., Cid Fernandes R., 2010, MNRAS, 403, 797 


\bibitem[\protect\citeauthoryear{Haines et al.}{2007}]{2007MNRAS.381....7H} 
Haines C.~P., Gargiulo A., La Barbera F., Mercurio A., Merluzzi P., 
Busarello G., 2007, MNRAS, 381, 7 

\bibitem[\protect\citeauthoryear{Heavens et 
al.}{2004}]{2004Natur.428..625H} Heavens A., Panter B., Jimenez R., Dunlop 
J., 2004, Nature, 428, 625 

\bibitem[\protect\citeauthoryear{Hughes 
\& Cortese}{2009}]{2009MNRAS.396L..41H} Hughes T.~M., Cortese L., 2009, MNRAS, 396, L41 



\bibitem[\protect\citeauthoryear{Iglesias-P{\'a}ramo et 
al.}{2013}]{2013A&A...553L...7I} Iglesias-P{\'a}ramo J., et al., 2013, A\&A, 553, L7 


\bibitem[\protect\citeauthoryear{Jimenez et 
al.}{2007}]{2007ApJ...669..947J} Jimenez R., Bernardi M., Haiman Z., Panter 
B., Heavens A.~F., 2007, ApJ, 669, 947 

\bibitem[\protect\citeauthoryear{Juneau et al.}{2014}]{2014ApJ...788...88J} 
Juneau S., et al., 2014, ApJ, 788, 88


\bibitem[Kauffmann et al.(2003)]{Kauffmannetal2003}
{Kauffmann}, G. and {Heckman}, T.~M. and {Tremonti}, C. et al., 2003, \mnras, 346, 1055

\bibitem[\protect\citeauthoryear{Kauffmann et 
al.}{2007}]{2007ApJS..173..357K} Kauffmann G., et al., 2007, ApJS, 173, 357 



\bibitem[\protect\citeauthoryear{Kehrig et 
al.}{2012}]{2012A&A...540A..11K} Kehrig C., et al., 2012, A\&A, 540, A11 

\bibitem[Kewley et al.(2001)]{Kewleyetal2001}
{Kewley}, L.~J. and {Dopita}, M.~A. and {Sutherland}, R.~S. et al., 2001, \apj, 556, 121

\bibitem[\protect\citeauthoryear{Kewley, Jansen, 
\& Geller}{2005}]{2005PASP..117..227K} Kewley L.~J., Jansen R.~A., Geller M.~J., 2005, PASP, 117, 227 



\bibitem[\protect\citeauthoryear{Kewley et al.}{2013}]{2013ApJ...774..100K} 
Kewley L.~J., Dopita M.~A., Leitherer C., Dav{\'e} R., Yuan T., Allen M., 
Groves B., Sutherland R., 2013, ApJ, 774, 100 



\bibitem[\protect\citeauthoryear{Kewley et al.}{2006}]{2006MNRAS.372..961K} 
Kewley L.~J., Groves B., Kauffmann G., Heckman T., 2006, MNRAS, 372, 961 

\bibitem[\protect\citeauthoryear{LaMassa et 
al.}{2013}]{2013ApJ...765L..33L} LaMassa S.~M., Heckman T.~M., Ptak A., 
Urry C.~M., 2013, ApJ, 765, L33 


\bibitem[\protect\citeauthoryear{Le Borgne et 
al.}{2003}]{2003A&A...402..433L} Le Borgne J.-F., et al., 2003, A\&A, 402, 433

\bibitem[\protect\citeauthoryear{Lee et al.}{2007}]{2007ApJ...663L..69L} 
Lee J.~H., Lee M.~G., Kim T., Hwang H.~S., Park C., Choi Y.-Y., 2007, ApJ, 
663, L69 

\bibitem[\protect\citeauthoryear{Lietzen et 
al.}{2011}]{2011A&A...535A..21L} Lietzen H., Hein{\"a}m{\"a}ki P., Nurmi P., Liivam{\"a}gi L.~J., Saar E., Tago E., Takalo L.~O., Einasto M., 2011, A\&A, 535, A21 



\bibitem[\protect\citeauthoryear{Martini}{2004}]{2004cbhg.symp..169M} 
Martini P., 2004, cbhg.symp, 169 

\bibitem[\protect\citeauthoryear{Mast et 
al.}{2014}]{2014A&A...561A.129M} Mast D., et al., 2014, A\&A, 561, A129 


\bibitem[\protect\citeauthoryear{Mateus et al.}{2006}]{2006MNRAS.370..721M} 
Mateus A., Sodr{\'e} L., Cid Fernandes R., Stasi{\'n}ska G., Schoenell W., 
Gomes J.~M., 2006, MNRAS, 370, 721



\bibitem[\protect\citeauthoryear{Ocvirk}{2010}]{2010ApJ...709...88O} Ocvirk 
P., 2010, ApJ, 709, 88 

\bibitem[\protect\citeauthoryear{Panter, Heavens, \& Jimenez}{2004}]{2004MNRAS.355..764P} Panter B., Heavens A.~F., Jimenez R., 2004, MNRAS, 355, 764


\bibitem[\protect\citeauthoryear{Papaderos et 
al.}{2013}]{2013A&A...555L...1P} Papaderos P., et al., 2013, A\&A, 555, L1 



\bibitem[\protect\citeauthoryear{Raimann \& Storchi-Bergmann}{2000}]{2000ASPC..221..213R} Raimann D., Storchi-Bergmann T., 2000, ASPC, 221, 213 


\bibitem[\protect\citeauthoryear{Rosario et 
al.}{2013}]{2013A&A...560A..72R} Rosario D.~J., et al., 2013, A\&A, 560, A72

\bibitem[\protect\citeauthoryear{Sarzi et al.}{2010}]{2010MNRAS.402.2187S} 
Sarzi M., et al., 2010, MNRAS, 402, 2187 

\bibitem[\protect\citeauthoryear{Schawinski et 
al.}{2014}]{2014arXiv1402.4814S} Schawinski K., et al., 2014, arXiv, 
arXiv:1402.4814 


\bibitem[\protect\citeauthoryear{Schawinski et 
al.}{2012}]{2012MNRAS.425L..61S} Schawinski K., Simmons B.~D., Urry C.~M., 
Treister E., Glikman E., 2012, MNRAS, 425, L61 

\bibitem[\protect\citeauthoryear{Schawinski et 
al.}{2010}]{2010ApJ...711..284S} Schawinski K., et al., 2010, ApJ, 711, 284 


\bibitem[\protect\citeauthoryear{Schawinski et 
al.}{2007}]{2007MNRAS.382.1415S} Schawinski K., Thomas D., Sarzi M., 
Maraston C., Kaviraj S., Joo S.-J., Yi S.~K., Silk J., 2007, MNRAS, 382, 
1415 

\bibitem[Scoville et al.(2006)]{Scovilleteal2006}
{Scoville}, N. and {Aussel}, H. and {Brusa}, M. et al., 2007, \apjs, 172, 1


\bibitem[\protect\citeauthoryear{Shlosman}{1990}]{1990NASCP3098..689S} 
Shlosman I., 1990, NASCP, 3098, 689 



\bibitem[\protect\citeauthoryear{Silk 
\& Rees}{1998}]{1998A&A...331L...1S} Silk J., Rees M.~J., 1998, A\&A, 331, L1 

\bibitem[\protect\citeauthoryear{Silverman et 
al.}{2009}]{2009ApJ...695..171S} Silverman J.~D., et al., 2009, ApJ, 695, 
171 

\bibitem[\protect\citeauthoryear{Singh et 
al.}{2013}]{2013A&A...558A..43S} Singh R., et al., 2013, A\&A, 558, A43 

\bibitem[Stasinska et al.(2006)]{Stasinskaetal2006}
{Stasi{\'n}ska}, G. and {Cid Fernandes}, R. and {Mateus}, A. et al., 2006, \mnras, 371, 972

\bibitem[Stasinska et al.(2008)]{Stasinskaetal2008}
{Stasi{\'n}ska}, G. and {Vale Asari}, N. and {Cid Fernandes}, R. et al, 2008, \mnras, 391, L29

 
\bibitem[\protect\citeauthoryear{Storchi-Bergmann, Calzetti, \& 
Kinney}{1994}]{1994ApJ...429..572S} Storchi-Bergmann T., Calzetti D., 
Kinney A.~L., 1994, ApJ, 429, 572 


\bibitem[\protect\citeauthoryear{Storchi-Bergmann et 
al.}{1996}]{1996ApJ...472...83S} Storchi-Bergmann T., Rodriguez-Ardila A., 
Schmitt H.~R., Wilson A.~S., Baldwin J.~A., 1996, ApJ, 472, 83 

 
\bibitem[\protect\citeauthoryear{Strauss et 
al.}{2002}]{2002AJ....124.1810S} Strauss M.~A., et al., 2002, AJ, 124, 1810 





\bibitem[\protect\citeauthoryear{Vale Asari et 
al.}{2009}]{2009MNRAS.396L..71V} Vale Asari N., Stasi{\'n}ska G., Cid 
Fernandes R., Gomes J.~M., Schlickmann M., Mateus A., Schoenell W., 2009, 
MNRAS, 396, L71 



\bibitem[\protect\citeauthoryear{van Zee et 
al.}{1998}]{1998AJ....116.2805V} van Zee L., Salzer J.~J., Haynes M.~P. et al., 1998, AJ, 116, 2805 

\bibitem[\protect\citeauthoryear{Veilleux}{2001}]{2001sgnf.conf...88V} 
Veilleux S., 2001, sgnf.conf, 88 

\bibitem[Vitale et al.(2013)]{Vitaleetal2013}
{Vitale}, M. and {Mignoli}, M. and {Cimatti}, A. et al., 2013, arXiv, 1304.2776

\bibitem[\protect\citeauthoryear{York et al.}{2000}]{2000AJ....120.1579Y} 
York D.~G., et al., 2000, AJ, 120, 1579 

\end{thebibliography}
\end{document}